\documentstyle[aps,pre,epsf,rotate]{revtex}

\begin{document}
\draft


\title{
On the theory of light scattering in molecular liquids
}

\author{A.Latz, M.~Letz\\
Institut f\"ur Physik, Johannes-Gutenberg Universit\"at, 55099
Mainz, Germany 
} 

\date{\today}

\maketitle

\begin{abstract}
The theory of light scattering for a system of linear molecules with
anisotropic polarizabilities is considered. As a starting point for
our theory, we express the result
of a  scattering experiment in VV and VH symmetry as 
dynamic correlation functions of tensorial
densities $\rho_{lm}(q)$ with $l=0$ and $l=2$. $l$, $m$ denote indices
of spherical 
harmonics.  To account for all observed hydrodynamic singularities, 
a generalization of the theory of Schilling and
Scheidsteger \cite{schilling97} for these correlation functions is
presented, which is capable to describe the light scattering
experiments from the liquid regime to the glassy state. As a
microscopic theory it fulfills all sum rules contrary to
previous {\em phenomenological} theories. 
We emphasize the importance of the helicity index $m$ for the
microscopic theory
by showing, that only the existence of $m=1$ components  lead to the well
known Rytov dip in liquids and to the appearance of transversal sound
waves in VH symmetry in the deeply supercooled liquid and the
glass. Exact expressions for the phenomenological frequency
dependent rotation
translation coupling coefficients of previous theories  are derived.   
\end{abstract}
\vskip -0.5cm
\section{Introduction}
\label{sec:intro}
In 1957 Rytov \cite{rytov57} predicted, based on a macroscopic
phenomenological theory, that the depolarized light scattering should
show a dip at zero frequency. Later in the 60's it was indeed
found experimentally \cite{stegeman68} 
that the light scattering of a liquid in VH
geometry, where the plane of polarization of the incident light and
the scattered light are perpendicular to each other, shows a symmetric
doublet at $\pm$ 1 GHz, with a minimum at zero frequency.  However
Rytov's theory only showed to be in superficial agreement with the
experiments. 

In the following time a number of theoretical attempts have been made
to understand the spectra. In general the intensity of the scattered
light $I^{SI}$ is proportional to the correlation function of the
fluctuations in the components $\alpha_{SI} (\vec{q},t)$ of the total 
polarizability tensor 
and to the incident light $I^I$. 
 Scattering in VH geometry is 
of particular interest since the direct (VV or HH) contributions are
dominated by Brillouin peaks which correspond to propagating sound
waves. In an ideal depolarized spectrum the Brillouin lines will be
absent
and therefore further information of the low frequency dynamics of
the system can be obtained.

It is most commonly believed that in molecular liquids, where
anisotropic single molecule polarizability fluctuations are the main
cause of depolarized spectra, the dip is caused by the coupling
of molecular orientations to some part of the stress tensor. These
theories can all be understood in the framework of the Mori--Zwanzig
projection technique by using different variables for the projection
scheme. Some primary variable is picked which couples via a memory
kernel to other secondary variables. Until the end of the 70th however
the generalized viscosities which occur in the memory kernel where
assumed to be time independent.

In 1969 Volterra \cite{volterra69} used the orientations as a primary
variable to which the stress tensor couples which he believed to have -- due to
symmetry reasons -- a nonzero matrix element with the total
polarizability. Anderson and Pecora proposed in 1971 \cite{anderson71}
a theory which used only the symmetric part of the stress tensor as a
secondary variable. Keyes and Kivelson in 1972 \cite{keyes72} used the
total polarizability as a primary variable and the momentum density as
a secondary. Ailawadi et al. \cite{ailawadi72} in 1972 coupled the
spin angular momentum density to the asymmetric part of the stress
tensor.

At that time it became evident that when reaching towards the supercooled
regime additional features occurred e.g. additional site peaks or a
central peak which
could not be explained consistently by any of the above two variable
theories. Therefore a number of theories were developed with
additional variables (and an additional number of adjustable
parameters). Quentrec in 1976 \cite{quentrec76} used 
the whole second rank tensor of the orientations
and Chappell et al. in 1981 \cite{chappell81}
coupled the momentum density and an unspecified symmetric tensor to
the total polarizability. And also a four variable theory with 7
adjustable parameters was proposed \cite{chappell82}. 
Wang in 1980 \cite{wang80} was the first to introduce time dependent
generalized viscosities which can account for retardation effects in
the memory functions. An approach which got refined and improved later on
\cite{wang86,dreyfus99}. The authors of
\cite{wang86} and  \cite{dreyfus99} emphasize the 
importance of retardation - (memory -) effects in the coupling
coefficients between the rotational and translational degrees of
freedom. Since all the mentioned theories are phenomenological, it is
difficult to decide, which one will give  the "correct" description of the
spectra. As an important result of our theory, which we are going to
present,  we will show, how the
most complex of the mentioned phenomenological theories
\cite{dreyfus99} can be derived microscopically as a special
approximation to a more complete set of equations. Thereby microscopic
expressions for the phenomenological coupling constants in  
\cite{dreyfus99} are presented.  

All of the above theories have in common, that they deal with the
theory of the experimental technique (light scattering) and the
dynamic of the system, which is measured in light scattering, on equal 
footing. One important deficiency of the  
early theories, is a too simplistic treatment for the dynamic of the
liquid (e.g. assuming white noise spectra for the in general
frequency dependent generalized viscosities \cite{anderson71}),
although the phenomenology for the light scattering part is correct.
From a conceptual point of view, a clear distinction
between these two topics should be made. The question which quantity  is
measured,   can and should be answered first. In a second, and we would 
like to stress, {\em independent} step, a theory for the measured
quantity can be formulated.  This strategy was already pursued in the 
generalized hydrodynamic approach of \cite{franosch00}, where -
without specifying the scattering mechanism - a completely general
treatment of the effect of hydrodynamic singularities and the
influence of slow structural relaxations on light scattering spectra
was achieved,  by deriving
formally exact expressions for frequency dependent Pockels constants
and related Green Kubo relations.   

Also in  our approach the two mentioned questions are clearly separated. But
contrary to \cite{franosch00},  we will formulate a theory for a
specific scattering mechanism, to get some more insight in the
microscopic mechanisms. We will show in section \ref{sec:other}, that
the formal structure of the
theory is compatible with \cite{franosch00}.  
First we will derive the quantities, which are measured in an ideal  
light scattering
experiments at  linear molecules, where all interaction induced
effects can be neglected. Then,  we present a set of microscopic
equations, which in principal enable us to calculate  these quantities
for a supercooled liquid close to the glass transition. In a final step
we approximate our equations very drastically, keeping only the
necessary ingredients for a qualitatively correct description of the
light scattering experiments.  This step is only for pedagogical
reasons, to demonstrate the ability of the theory to reproduce light
scattering spectra close to the glass transition.

This paper is organized as follows: 
In section \ref{sec:sec3} we derive the
direct contribution to the light scattering spectrum of a liquid of linear
molecules. It will provide the starting point for the following 
theoretical considerations.    

In section \ref{sec:mmct} we present a straight forward generalization 
of the
molecular mode coupling theory \cite{schilling97} for linear molecules, 
which in addition to
the orientational and translational degrees of freedom also describes
the coupling to transversal current fluctuations. 
In section \ref{sec:glasd} we formulate a restricted {\em not} self consistent
theory, which still
contains all hydrodynamic modes and the most important rotational
degrees of freedom.  As a further, very drastic approximation, we set up 
a simple Maxwell theory to demonstrate that
all qualitative features, observed in light scattering experiments in
supercooled liquids are reproduced already within the restricted
theory, by changing only the 
time scale of the structural relaxation.   In section \ref{sec:other}
the relation to other light scattering theories is discussed. The 
phenomenological equations of \cite{dreyfus99} are derived within our
theory and shown to be a special case of the restricted theory.

\section{Light scattering from molecular systems}
\label{sec:sec3}
Following \cite{madden91} the total polarizability of a molecular
liquid can be expressed as
a sum of single particle and interaction induced many particle
contributions.
\begin{eqnarray}
\mbox{\boldmath $\alpha$}(r-r') E^{ext}(r') &=& \sum_i
\mbox{\boldmath $\alpha$}^i \delta(r-r^i) \; \delta(r'-r^i) E^{int}(r') 
\nonumber \\
&&+ \sum_{ij}
\mbox{\boldmath $\alpha$}^{ij} \delta(r-r^i) \; \delta(r'-r^j) E^{int}(r') 
\nonumber \\
&&+  \sum_{ijk} \mbox{\boldmath $\alpha$}^{ijk} ...
\label{eq:polges}
\end{eqnarray}
Where the superscript ${ext}$ denotes the external electric field and
${int}$ denotes an effective internal field.
In this work we are concentrating on the direct contribution for
linear molecules only. In a dense liquid other scattering mechanisms
like direct and indirect dipole induced dipole scattering mechanism
and collision induced scattering mechanisms  are also present, in
general.
For CS$_2$ e.g. it has been argued by Madden and Tildesley \cite{madden85}
that the interaction induced mechanism is the dominant one (although
this result has been questioned \cite{murry99}).  For Salol 
it was found \cite{cummins96,cummins97err}, that the depolarized light
scattering is 
dominated by scattering at orientational fluctuations. In
\cite{cummins96,cummins97err} more examples are discussed. 
A theory for depolarized DID spectra in simple liquids was presented
in \cite{fuchs91}. A generalization of this work to a hydrodynamic
theory of of light scattering, which incorporates in principle all
possible scattering mechanisms was developed in \cite{franosch00}.  
Here we want to concentrate on the molecular origin of the depolarized 
light scattering for a system of linear molecules. A generalization to
non linear molecules is straight forward, but to avoid unnecessary
complexity, we want to restrict ourselves to linear molecules. The
numerical study of water \cite{fabbian99,theis00} has shown, that     
even the dynamics of this non linear molecule 
can be reasonably well reproduced  by modeling it as as a linear molecule. 
Only  direct
contribution of the orientations to the spectra are discussed. All
interaction induced contributions are neglected. 

For linear molecules the polarization tensor of a single molecule 
can be written in the form

\begin{equation} \label{alpha}
\mbox{\boldmath $\alpha$}^{(i)} = a {\bf I} + \frac{g}{3} (3
\hat{n}^{(i)}  \hat{n}^{(i)} - {\bf I}) 
\end{equation}
Here $\hat{n}^{(i)}$ is a unit vector in the direction of the principal
axis of the molecule $i$. $\bf I $  is the identity  matrix.  The numbers
$a$ and $g$ are the mean polarizability and the anisotropy of the
polarization respectively.

 The collective wave-vector dependent
fluctuations of the total polarizability tensor $\mbox{\boldmath
  $\alpha$} (\vec{q},t)$ are defined by
\begin{equation} \label{polcollective}
\delta \alpha_{SI} (\vec{q},t) = \sum_{i=1}^N \delta
\alpha^{(i)}_{SI}(t) \exp (i \vec{q} \vec{r_i(t)}).   
\end{equation}
Here the indices $S$,  $I$ denote the direction of the scattered and
incident light, respectively. The fluctuation of the single molecule 
polarizability $\delta \alpha^{(i)}_{SI}(t)$ is completely
due to rotations of the molecule. 

\begin{figure}
\unitlength1cm
\epsfxsize=10cm
\begin{picture}(9,11)
\put(-0.5,0){\rotate[r]{\epsffile{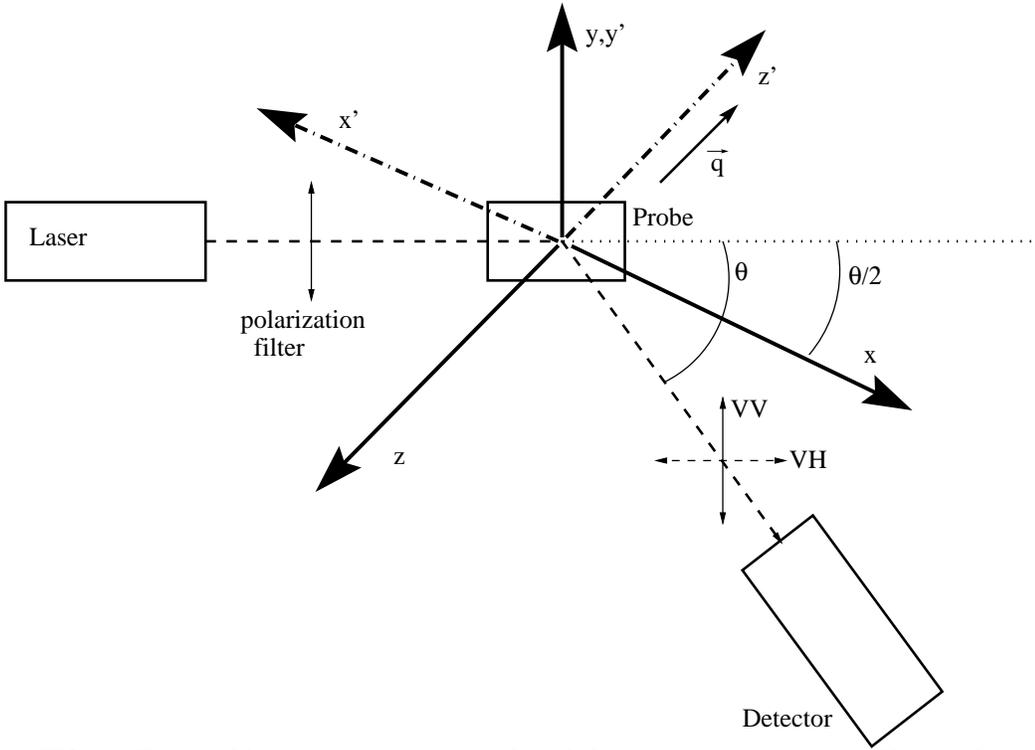}}}
\end{picture}
\caption{The possible scattering geometries for a light-scattering
experiment. The coordinate system which is denoted with
$(x,y,z)$ is the common coordinate system used in light scattering
theories. The coordinate system which is denoted with $(x',y',z')$ is
the $q$ - frame usually used in theoretical descriptions for the dynamics
of liquids where the z-axis points along the q-vector (arrow) which is probed.
}
\label{fig:geom}
\end{figure}

The light scattering intensity is then given by 
\begin{eqnarray}
I_{SI}(q,\omega) &\sim& FT(\langle
\alpha_{SI}^*(\vec{q},t) 
 \alpha_{SI}(\vec{q},0) 
\rangle ) \nonumber \\
&=& 
\int_{-\infty}^{\infty} dt e^{i\omega t} 
\nonumber \\ 
&& \mbox{\hspace*{0.5cm}}
\langle
\sum_{i,j} {\alpha_{SI}^{(i)}}^*(t)e^{-iqr_i(t)} 
\alpha_{SI}^{(j)}(0)e^{iqr_j(0)} 
\rangle
\end{eqnarray}
where the $\langle \rangle$ brackets denote a thermal expectation
value. 
Usually the convention is
used that the xz-plane is the scattering plane and the scattering
vector is anti-parallel to the z-axis (see Fig. \ref{fig:geom}).  In
this geometry the 
depolarized scattering in VV and VH geometry, respectively, read : 
\begin{eqnarray} \label{spectra}
I^{VV}(q,\omega) &=& I_I FT (\langle \delta
  \alpha_{yy}(q,t)^* \delta \alpha_{yy}(q,0) \rangle)(\omega)
\label{spectravv} \\
I^{VH}(q,\omega) &=& I_I FT \left ( \langle
\delta \alpha_{yx}^*(q,t) 
\delta \alpha_{yx}(q,0) 
\rangle \sin^2(\Theta/2) \right . \nonumber \\
&& + \left . \langle
\delta \alpha_{yz}^*(q,t) 
\delta \alpha_{yz}(q,0) 
\rangle \cos^2(\Theta/2) 
\right )(\omega) \label{spectravh}
\end{eqnarray}
where $\Theta$ is the scattering angle, $q =|\vec{q}|$ and $FT$
denotes the Fourier
transform.
For linear molecules, the specified
scattering frame is equivalent to the so called $q$ -frame, where the
z - axis is parallel to the scattering vector $\vec{q}$, since the
polarizability tensor is a spherical tensor of rank two, where only
$l=0$ and $l=2$ components appear (see appendix \ref{app:a}). 
These are invariant under
reflection $\vec{q} \to -\vec{q}$. 
The explicit form of {\boldmath $\alpha$} in the $q$ - frame is
derived in appendix \ref{app:a}.  With (\ref{eq:polmat}) we 
arrive at the result 
\begin{eqnarray} 
I^{VV}(q,\omega) &\sim&  a^2 {S''}_{00}^0(q,\omega) +  g^2 \frac{4 \pi}{15}
({S''}_{22}^2(q,\omega) +   \frac{1}{3} {S''}_{22}^0(q,\omega)) - 
a g \frac{4}{3} \sqrt{\frac{\pi}{5}} {S''}_{20}^0(q,\omega) \label{eq:vv}\\
I^{VH}(q,\omega) &\sim& g^2 \frac{4 \pi}{15}( \sin^2(\Theta/2)
{S''}_{22}^2(q,\omega) + \cos^2(\Theta/2) \label{eq:hv}
{S''}_{22}^1(q,\omega)) 
\end{eqnarray}
In the derivation of Eqs. (\ref{eq:vv}) and (\ref{eq:hv}) we used that
in the q -frame, 
the dynamic structure factors $S_{ll'}^{mm'}(q,t)$ are diagonal
with respect to $m,m'$  and
that for linear molecules $S_{ll'}^m(q,t) =
S_{ll'}^{-m}(q,t)$. The
contribution $\sim {S''}_{20}^0(q,\omega) $ can be neglected for light
scattering experiments, since it is of relative order $q^2$, compared
to ${S''}_{22}^0(q,t)$ and ${S''}_{00}^0(q,t)$. 

Similar equations were
already derived in \cite{berne76} for the special case of a diluted
gas, where translational and rotational motion can  be factorized.
In the case of dense liquids, the form  Eqs. (\ref{eq:vv}) and
(\ref{eq:hv})  have to be used. 
For a system of linear molecules with only direct but
anisotropic scattering
mechanism, they provide a complete and exact description of the observed
light scattering spectra. Since we are interested in light scattering
experiments in supercooled liquids, we are going to present a
microscopic theory
for the generalized dynamic structure factors $S_{ll'}^m(q,t)$ which
contains all observed features  of light scattering 
experiments close to the glass transition. Already from the general
form of (\ref{eq:vv})  and (\ref{eq:hv}) some general conclusions can be
drawn. 

First, due to the scattering angle $\Theta$ in the VH geometry a
backscattering geometry only observes the $m=2$ component of the
dynamic density correlation function whereas a 90 degree scattering
angle probes a mixture of $m=1$ and $m=2$ components.
In the following sections we show that the $S_{22}^0(q,\omega)$ and 
$S_{22}^1(q,\omega)$
component couple to the longitudinal and transverse sound mode,
respectively, while the
$S_{22}^2(q,\omega)$ component does not couple to any hydrodynamic
mode. Therefore it is possible to replace $S_{22}^2(q,\omega)$ by its
value at $q=0$, but not the correlators with $m=0,1$. The
different behavior of the correlators with different m is due to a
dynamic breaking of the rotational
invariance on the spatial scale of the light scattering experiments,
caused by the existence of hydrodynamic singularities. The
phenomenological form for the light scattering  spectra used
in the literature can be
recovered by  rewriting
Eqs. (\ref{eq:vv}) and (\ref{eq:hv}) in the form 
\begin{eqnarray} 
I^{VV}(q,\omega) &\sim&  a^2 {S''}_{00}^0(q,\omega) +  g^2 \frac{4 \pi}{15}
(\frac{4}{3}{S''}_{22}^0(q=0,\omega) +   \frac{1}{3} ({S''}_{22}^0(q,\omega)
- {S''}_{22}^0(q=0,\omega)))\label{eq:vv_phen}\\
I^{VH}(q,\omega) &\sim& g^2 \frac{4 \pi}{15}(
{S''}_{22}^0(q=0,\omega) + \cos^2(\Theta/2) \label{eq:hv_phen}
({S''}_{22}^1(q,\omega)-{S''}_{22}^0(q=0,\omega))) 
\end{eqnarray}
Here we have used that for $q=0$ due to rotational symmetry all 
correlation
functions with different helicity $m$ but the same $l$ are equal and
are therefore replaced with $m=0$. In addition 
we have neglected the off diagonal contribution
$S_{20}^0(q,\omega)$. The last two terms in Eqs. (\ref{eq:vv_phen})
and (\ref{eq:hv_phen}) contain hydrodynamic poles and thus 
essential for the understanding of light scattering experiments.
For theoretical considerations it is more convenient to use Eqs. (\ref{eq:vv})
and (\ref{eq:hv}).   Therefore we continue in this work using the
representation in spherical harmonics.

Second, the spectra fulfill sum rules. The total intensities
$I^{SI}_{tot} = \int d 
\omega I^{SI}_{tot} (\omega)$ for $q \to 0$ are derived from
Eqs. (\ref{eq:vv}) and (\ref{eq:hv}) 
\begin{eqnarray}
I^{VV}_{tot}(q=0,t=0) &\sim&  a^2 {S''}_{00}^0(q=0,t=0) +  g^2 \frac{4 \pi}{15}
\; \; \frac{4}{3} {S''}_{22}^0(q=0,t=0)  \label{ivvtot}\\
I^{VH}_{tot}(q=0,t=0) &\sim& g^2 \frac{4 \pi}{15}
{S''}_{22}^0(q=0,t=0)  \label{ivhtot} \;\;. 
\end{eqnarray}
Here we used, that the static structure factor $S_{ll'}^m(q=0,t=0)$ are
independent of $m$ and diagonal in $l,l'$.  Therefore the anisotropic part 
of $I^{VV}_{tot}$ is exactly equal to $\frac{4}{3}  I^{VH}_{tot}$. 
  
\begin{figure}
\unitlength1cm
\epsfxsize=10cm
\begin{picture}(9,8)
\put(-0.5,-1){\rotate[r]{\epsffile{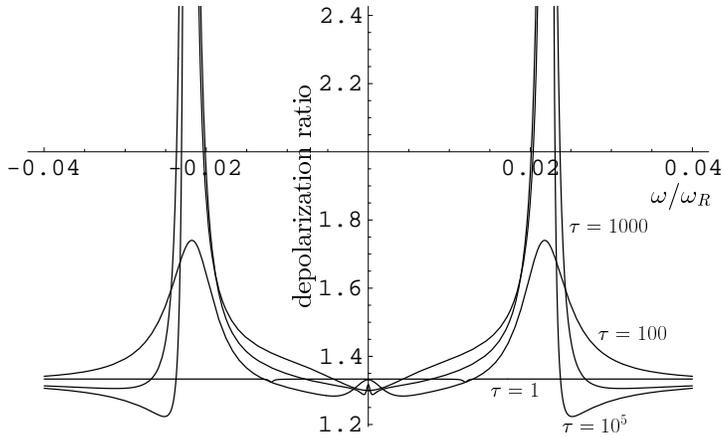}}}
\end{picture}
\caption{The depolarization ratio for backscattering geometry
{\protect $D(q,\omega) = \left( ({S''}^2_{22}(q,\omega) + 1/3
{S''}^0_{22}(q,\omega)\right ) /
{S''}^2_{22}(q,\omega)$ }
for some arbitrary chosen parameters. 
In units of the rotational frequency $\omega^R\equiv 1$, the other parameters
where chosen as $c_{\parallel}=0.6$, $G_s=1$, $\nu_R=5$, $K_l=1$,
$K_R=1$, $K_{lR}=K_{SR}=1/2$. The values of the 
$\alpha$-relaxation time $\tau$ 
were set identical for all
components with $\tau=\tau_l=\tau_R=\tau_{SR}=\tau_{lR}=\tau_S \in
\{1,\;0.01,\;0.001,\; 
10^{-5} \}$.
}
\label{fig:dep}
\end{figure}

\begin{figure}
\unitlength1cm
\epsfxsize=10cm
\begin{picture}(9,8)
\put(-0.5,-1){\rotate[r]{\epsffile{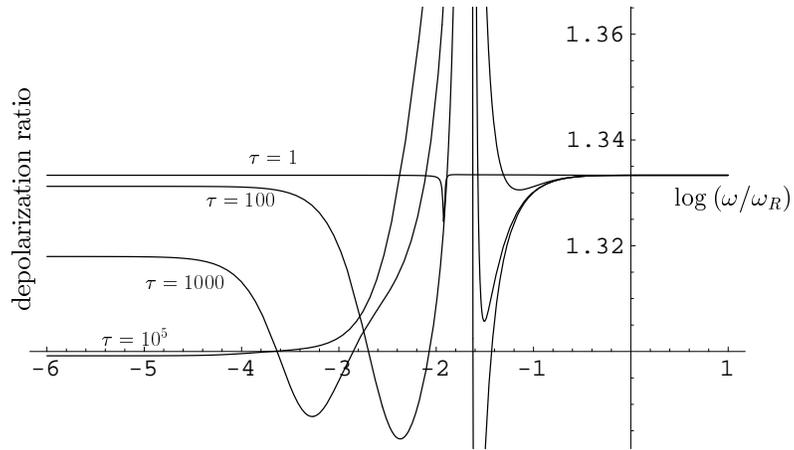}}}
\end{picture}
\caption{
The depolarization ratio for the same parameters as in
Fig. \protect\ref{fig:dep} on a logarithmic frequency scale.
}
\label{fig:deplog}
\end{figure}

\begin{figure}
\unitlength1cm
\epsfxsize=10cm
\begin{picture}(9,8)
\put(-0.5,-1){\rotate[r]{\epsffile{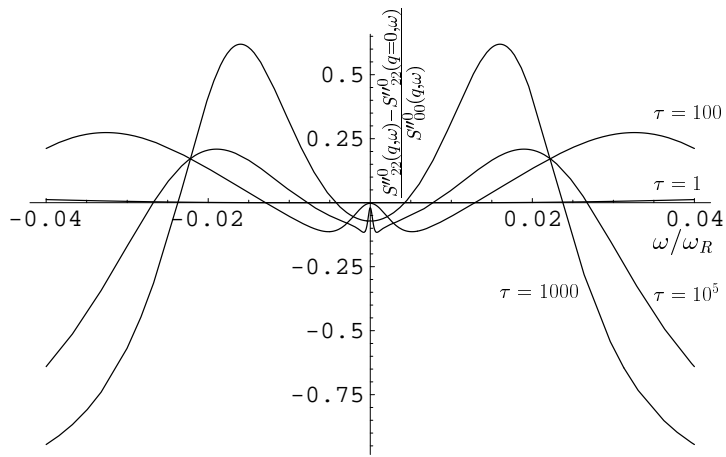}}}
\end{picture}
\caption{The relation discussed in Eq. (\protect{\ref{eq:depref}}) is
plotted on a linear scale.
}
\label{fig:refrep}
\end{figure}

\begin{figure}
\unitlength1cm
\epsfxsize=10cm
\begin{picture}(9,8)
\put(-0.5,-1){\rotate[r]{\epsffile{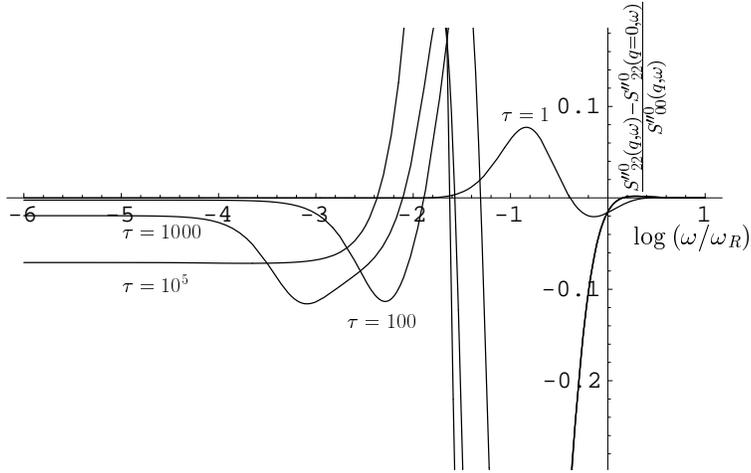}}}
\end{picture}
\caption{The relation discussed in Eq. (\protect{\ref{eq:depref}}) is
plotted on a logarithmic scale.
}
\label{fig:refantwlog}
\end{figure}

Third, we note the well established fact, that  the often used
procedure  to obtain the
isotropic scattering contribution (e.g.  in \cite{fioretto99}, see
also \cite{berne76}) by
simply subtracting $\frac{4}{3} I^{VH}$ 
from $I^{VV}(\omega)$ is in general not exact, since there are big
qualitative  
differences between the correlation functions $S_{22}^m (q,\omega)$
for different $m$, due to the coupling of translational and rotational 
motion. In Appendix 7B of \cite{berne76} the mentioned relation
between $I^{VV}(\omega)$ and $I^{VH}(\omega)$ could be derived by
explicitly assuming, that rotational and translational motion are
independent. But it is clear, that this cannot hold in dense liquids,
where the rotation of a molecule can easily cause the build up of local
stress via interaction with its neighbors. This stress can then be
released by a center of mass motion of neighboring molecules.
Instead of obtaining only the isotropic contribution the mentioned
subtraction method will yield the following expression. 
\begin{eqnarray} 
\label{eq:depref}
\lefteqn{I^{VV}(q,\omega) - \frac{4}{3}
I^{VH}(q,\omega)} \nonumber \\
 &\sim&  a^2
{S''}_{00}^0(q,\omega) +\label{eq:subtract} \\
&&  g^2 \frac{4 \pi}{45} ({S''}_{22}^0(q,\omega)
- {S''}_{22}^0(q=0,\omega) - 4 \cos^2(\Theta/2) 
({S''}_{22}^1(q,\omega)-{S''}_{22}^0(q=0,\omega))) \nonumber
\end{eqnarray}
Eq. (\ref{eq:subtract}) shows, that even in back scattering geometry
deviations from the purely isotropic scattering are to be expected.   
To demonstrate this point in more detail, we plot 
in figure \ref{fig:dep} the value of the
depolarization ratio in backscattering geometry ($\Theta = \pi$) 
for a simple model which we derive from our
equations (see section \ref{sec:glasd}). 
To further demonstrate the result of Eq. (\ref{eq:depref}) we have
(using the same model discussed in section \ref{sec:glasd}) further
plotted the quantity
${S''}_{22}^0(q,\omega)-{S''}_{22}^0(q=0,\omega))/{S''}_{00}^0(q,\omega)$.
This is 
done on a linear frequency scale in Fig. \ref{fig:refrep} and on a logarithmic
scale in Fig. \ref{fig:refantwlog}. It can be clearly seen that deep
in the liquid phase ($\tau=1$) it is practically zero whereas strong
deviations occur esspecially around the Brillouin lines when
supercooling the liquid.
From Eqs.  (\ref{eq:vv_phen})
and (\ref{eq:hv_phen}) it follows that the depolarization ratio in
general is given by 
\begin{eqnarray}
\lefteqn{\frac{
{S''}^2_{22}(q,\omega) + \frac{1}{3} {S''}^0_{22}(q,\omega)}
{\sin^2(\Theta/2) {S''}^2_{22}(q,\omega) +\cos^2(\Theta/2)
{S''}^1_{22}(q,\omega)}} \nonumber \\ &=&
\frac{4}{3} + \frac{1}{3} 
\frac{{S''}_{22}^0(q,\omega) -
{S''}_{22}^0(q=0,\omega)}{{S''}_{22}^0(q=0,\omega) + \cos^2(\Theta/2)
({S''}_{22}^1(q,\omega) -{S''}_{22}^0(q=0,\omega))}  
\end{eqnarray}
In backscattering geometry ($\Theta=\pi$)  it is nearly constant in
the liquid and given by $4/3$ whereas strong
deviations occur around the Brillouin line when super-cooling the
liquid. These deviations are expected to be even bigger when the
experiment is performed not in backscattering geometry since in this
case the $m=1$ component which couples to the transverse phonon will
also contribute. Fig. \ref{fig:deplog} shows the depolarization ratio
on a logarithmic frequency scale.
The deviations of the depolarization ratio from $4/3$ are also
seen in the experiment and are discussed in \cite{cummins96}. 
Therefore we have proven that the quantity $I_{VV}- 4/3 I_{VH}$ is
{\it not} proportional to ${S^0_{00}}''(q,\omega)$, especially not
around the Brillouin lines. We are aware that this is a severe
difficulty in interpreting light--scattering experiments and want to
point out that in the low frequency regime it is usually not applied
anyways since $I_{VH}$ becomes very weak.

\section{The equation of motion of linear molecules} 
\label{sec:mmct}

After having an exact expression for the scattered intensities (see
Eqs. (\ref{eq:vv})  and
(\ref{eq:hv})), we turn in this section to the second and independent
step in order to arrive at a microscopic understanding of light
scattering spectra. This second step is the discussion of a
microscopic theory of a molecular liquid.
The mode coupling theory of the glass transition has by now proven to
be a very successful microscopic theory to describe the dynamics of supercooled
liquids close to the glass transition (for a recent review see
e.g. \cite{goetze99r,cummins99}). In its early  version
\cite{bengtzelius84} only simple liquids consisting of spherical
molecules were described microscopically. Due to its successful
application also to experiments and simulations at arbitrary molecules 
it became necessary to develop generalizations to molecular liquids
\cite{schilling97,franosch97,kawasaki97,chong98,fabbian99}. Also
aspects of these  generalized theories were successfully tested in
simulations  
\cite{winkler00,fabbian00,theis00}. The main quantity, which is
studied in \cite{schilling97} is the coherent dynamic structure factor 
$S_{ll'}^m(q,t)$ for linear molecules.  The theory describes on a
microscopic basis  the coupling of translational and orientational
degrees of freedom, but does not take into account the coupling to
transversal currents.   
In \cite{fabbian99} the theory
for the dynamic structure factor $S_{ll';nn'}^{mm'}(\vec{q},t)$ of
arbitrary rigid molecules was  derived. This theory deals also with
the transversal currents, but is at present difficult to reduce to linear 
molecules.  Since all theories of light scattering agree in the
importance of transversal current fluctuations, it will be necessary
to formulate
a theory for linear molecules, which also contains coupling to
transversal currents. Fortunately this theory is a straight forward
generalization of the  theory in \cite{schilling97}. We do not take
into account energy fluctuations in our derivation, which were  necessary
to describe the Rayleigh peak i.e. the influence of heat diffusion on
the light scattering spectrum. We do not expect important changes for the
discussion of the light scattering spectrum, by neglecting the
influence of heat fluctuations. For a discussion of the interplay of
structural relaxation and heat diffusion see \cite{latz89,franosch00}.

The MMCT is derived within a Mori Zwanzig projection operator formalism 
for tensorial densities $\rho_{lm}(\vec{q},t)$    
\begin{eqnarray} \label{rholm}
\rho_{lm}(\vec{q},t) &=& \sqrt{4 \pi} i^l \sum_{i=1}^N 
Y_{lm}(\Omega_i(t)) \; e^{i\vec{q} \vec{r}_i(t)}
\end{eqnarray}
with $Y_{lm}(\Omega)$ being the spherical harmonics 
and tensorial currents $\{\vec{j}^{\alpha}\}_{lm}(\vec{q},t)$.  The index
$\alpha =T,R$ denote translational currents and rotational currents,
respectively. For linear molecules it is not necessary to use the 
orientational current as a vectorial quantity.
The only quantity which will appear in the theory is 
\begin{equation}\label{jrotlm}
\{j^R_0\}_{lm}(\vec{q},t) = \frac{\sqrt{4 \pi}}{\sqrt{l(l+1)}} i^l \sum_{i=1}^N 
(\vec{\omega} \vec{L}) Y_{lm}(\Omega_i(t)) \; e^{i\vec{q} \vec{r}_i(t)}.
\end{equation}
Here $\vec{L}$ is the angular momentum operator and $\vec{\omega}$ is
the angular velocity. $\{j^R_0\}_{lm}(\vec{q},t)$ with $m \in \{-l,-l+1,
\ldots l-1,l\}$ are the components of an  irreducible
spherical tensor of rank $l$.  

To be able to describe light scattering spectra we will need all
components of the translational current fluctuations
$\{j^T_\mu\}_{lm}(\vec{q},t)$, $\mu \in \{-1,0,1\}$,  
not only its projection along $\vec{q}$ as in \cite{schilling97}. Here 
we used spherical components $\mu$ instead of Cartesian components
$\{x,y,z\}$.  

\begin{equation}\label{jtranslm}
\{j^T_\mu\}_{lm}(\vec{q},t) = i^l \sqrt{4 \pi} \sum_{i=1}^N 
v_\mu Y_{lm}(\Omega_i(t)) \; e^{i\vec{q} \vec{r}_i(t)}.
\end{equation}

with 
\begin{eqnarray}
\{j^T_{\pm 1}\} _{lm}(\vec{q},t) &=& \frac{-1}{\sqrt{2}}
(\pm \{j^T_x\}_{lm}(\vec{q},t) + 
i \{j^T_y\}_{lm}(\vec{q},t))\nonumber \\ 
\{j^T_0\}_{lm}(\vec{q},t) &=& \{j^T_z\}_{lm}(\vec{q},t)
\end{eqnarray}
The translational currents $\{j^T_\mu\}_{lm}(\vec{q},t)$ can be written
as a sum of components of an irreducible spherical tensor
$\{j^{T}\}^{irr}_{\hat{l} \hat{m}} (\vec{q},t)$ 
\begin{equation} \label{pm1}
\{j^T_\mu\}_{lm}(\vec{q},t) = \sum_{\hat{l},\hat{m}} C(1l\hat{l};\mu m
\hat{m}) \{j^{T}\}^{irr}_{\hat{l} \hat{m}} (\vec{q},t)
\end{equation}
with $|l-1| \le \hat{l} \le l+1$ and $\hat{m} = m+\mu$, due to the
properties of the Clebsch Gordan coefficients $C(l_1l_2l;m_1m_2m)$.
As an important special case we note, that the transversal
center of mass currents  $(j^T_{\pm 1})_{00}(\vec{q},t)$ have
irreducible spherical components $\{j^{T}\}^{irr}_{1 \pm 1}(\vec{q},t)$. 

The currents and the densities fulfill a continuity 
equation 
\begin{equation}\label{cont}
\frac{d}{dt} \rho_{lm}(q,t) = \sum_{\mu,\alpha} i (-1)^\mu q^\alpha_{-\mu}(l)
\{j^\alpha_\mu\}_{lm} (q,t) 
\end{equation}
with 
\begin{equation}\label{eq:qmu}
q^\alpha_{\mu}(l) = \left\{ \begin{array}{c@{\quad \mbox{for} \;\;}c}
    q_{\mu}& 
    \alpha = T\\
\sqrt{l(l+1)} &
    \alpha = R \end{array} \right.
\end{equation}

With the help of Mori - Zwanzig  projection operator techniques
\cite{forster83,schilling97}, it is possible to derive a formally exact set of
equations for the correlation functions $S_{ll'}^{mm'}(\vec{q},t)$  of  the 
tensorial densities $\rho_{lm}(\vec{q},t)$. For simplicity we write the
equation in the $q$ -  frame i.e. ($\vec{q} = q(0,0,1)$). 
For a short ranged potential every
correlation function $(A_{lm}(q,t)|B_{l'm'}(q)) = \langle
A_{lm}(q,t)^* \; B_{l'm'}(q) \rangle $ of two  components
$A_{lm}(q,t), B_{l'm'}(q)$ of  
 irreducible spherical tensors fulfill the relations
\begin{eqnarray} \label{diag}
(A_{lm}(q,t)|B_{l'm'}(q)) &=& \delta_{mm'} (A_{lm}(q,t)|B_{l'm}(q))\\
(A_{lm}(q,t)|B_{l'm}(q)) & \sim&  q^{|l-l'|} \;\; \mbox{for $q \to
0$} \nonumber . 
\end{eqnarray}
The first line or (\ref{diag}) is just due to the fact that under a
rotation around the z - axis, 
$(A_{lm}(q,t)|B_{l'm'}(q))$ will transform into $e^{i (m'-m)\phi}
(A_{lm}(q,t)|B_{l'm'}(q))$. Since the correlation function has to be
invariant under this operation, the relation (\ref{diag}) follows. The 
second line is a consequence of global rotational invariance of the
system of linear molecules. 
In addition, the correlation function $(A_{lm}(q,t)|B_{l'm}(q))$ is
independent of $m$ for $q=0$, if there are no long range correlations, which
destroy the global rotational invariance. For correlation functions
which behave regular at $q=0$ i.e. which do not contain any
hydrodynamic poles, we are allowed to replace them for small $q$ by
their value at $q=0$ plus corrections. If there are hydrodynamic
poles, the differences between different $m$ can be crucial e.g. we
will see, that the dynamic correlation function $S_{22}^m(q,z)$
contains for $m=1$ even at small values of q couplings to transversal  current
fluctuations, where the one for $m=2$ behaves regular at $q
\to 0$.

The dynamic correlation function $S_{ll'}^{mm'}(q,t)$ is a real
quantity \cite{schilling97}. Since we will discuss in the
following part of the paper also current--current correlation
functions we use the notation 
$ S_{ll'}^m(q,t) \equiv  (\underline{\underline{\phi}}_{\rho
\rho})_{ll'}^{mm}(q,t)$
The equation for $\underline{\underline{\phi}}_{\rho
\rho}(q,t)$ is therefore 
\begin{eqnarray}
\label{eq:mctanf}
\frac{\partial}{\partial t} 
\underline{\underline{\phi}}_{\rho \rho}(q,t)  
&=& -i \; \; \underline{\underline{\vec{q}}}  
\; \underline{\underline{\phi}}_{\vec{j}
  \rho}(q,t)  
\equiv -i \; \;  \; \sum_{\alpha}
\underline{\underline{q}}_0^\alpha  \;
\underline{\underline{\phi}}_{j_0^\alpha
\rho}(q,t) 
\nonumber \\
 \frac{\partial}{\partial t} 
\underline{\underline{\phi}}_{\vec{j} \rho}(q,t) &=&
 - i \;  
 \underline{\underline{\Gamma}}_{\vec{j} 
\rho}(q)\; 
\underline{\underline{\phi}}_{\rho \rho}(q,t)
\nonumber \\ 
&& \mbox{\hspace*{-2.0cm}}
-  \int_0^t \; dt'\;
\underline{\underline{M}}_{\vec{j} \vec{j}'}(q,t-t') \;
\underline{\underline{\phi}}_{\vec{j}' \rho}(q,t) \nonumber \\
\frac{\partial}{\partial t} 
\underline{\underline{\phi}}_{\rho \vec{j}'}(q,t)  
&=& i \; \underline{\underline{\vec{q}}}  \; 
\underline{\underline{\phi}}_{\vec{j}
\vec{j}'}(q,t)  
\equiv i \;  \;  \; \sum_{\alpha}
\underline{\underline{q}}_0^\alpha  \;
\underline{\underline{\phi}}_{j_0^\alpha \vec{j}'}(q,t)
\nonumber \\
 \frac{\partial}{\partial t} 
\underline{\underline{\phi}}_{\vec{j} \vec{j}'}(q,t) &=&
 - i \; 
 \underline{\underline{\Gamma}}_{\vec{j} 
\rho}(q)\; 
\underline{\underline{\phi}}_{\rho \vec{j}'}(q,t)
\nonumber \\ 
&& \mbox{\hspace*{-2.0cm}}
-  \int_0^t \; dt'\;
\underline{\underline{M}}_{\vec{j} \vec{j}''}(q,t-t') \;
\underline{\underline{\phi}}_{\vec{j}'' \vec{j}'}(q,t) 
\end{eqnarray}
where we use the short--hand notation 
$\vec{j} =\{j^{\alpha}_{\mu}\}_{lm} $, $\vec{j}' =
 \{j^{\alpha'}_{\mu'}\}_{l'm'}$ and    
$\underline{\underline{\vec{q}}}= \delta_{ll'} (-1)^\mu q_{-\mu}^\alpha(l)$,
 $\underline{\underline{\phi}}_{j_\mu^\alpha \rho}(q,t) =
\delta_{m \; m'+\mu} (\{j^\alpha_\mu\}_{lm}(q,t)| 
\rho_{lm'}(q,0)) $, $\underline{\underline{\phi}}_{\rho
j_\mu^\alpha}(q,t) = 
\underline{\underline{\phi}}^+_{j_\mu^\alpha \rho}(q,t)$ and
$\underline{\underline{\phi}}_{j_\mu^\alpha 
j_{\mu'}^{\alpha'}}(q,t) = \delta_{\mu+m,\mu'+m'} (\{j^\alpha_\mu\}_{lm}(q,t)|
\{j^{\alpha'}_{\mu'}\}_{l'm'}(q,0)))$ 
are the current--density, density--current and current--current
correlation functions, respectively. The
matrix ${\underline{\underline{\Gamma}}}_{\vec{j} \rho}(q) =
\delta_{\mu,0} 
\underline{\underline{q}}_0^\alpha \frac{k_BT}{\Theta_{\alpha}} ({\bf
S}^{-1}(q,m))$ is determined by the static molecular correlators. 
$\Theta_{T}$ and 
$\Theta_{R}$ are the mass and inertia,
respectively. 
The products 
$\underline{\underline{q}}_0^R
{\underline{\underline{\Gamma}}}_{j_0^R \rho}(q)$ and
$\underline{\underline{q}}_0^T 
{\underline{\underline{\Gamma}}}_{j_0^T \rho}(q)$ are the matrix of
rotational and translational microscopic frequencies, respectively. 
The memory matrix $\underline{\underline{M}}(q;t) = {\bf M}_{\mu
  \mu'}^{\alpha \alpha'} 
= \{ M ^{\alpha,\alpha'}_{\mu, \mu'}\}^{l,l'}_{m,m'} (q;t) =
(Q {\cal L} (\{j_\mu^\alpha(q)\}|R'(t)|Q {\cal L} (\{j_{\mu'}^{\alpha'}(q)\})
\frac{\Theta_{\alpha'}}{k_BT} $ is a frequency dependent damping matrix, 
${\cal L}$ is the
Liouville operator, $Q$ is the
projection operator perpendicular to the density- and current
fluctuations (see Eqs. (\ref{rholm}), (\ref{jrotlm}) and
(\ref{jtranslm})). $R'(t)$ is a 
reduced time translation operator $R'(t) = Q e^{i
Q{\cal L}Qt}$ \cite{forster83}. 
The memory matrix is {\em not} 
diagonal in $m$ and $m'$, contrary to the one appearing  in 
\cite{schilling97}. This is due to the fact, that the currents for
$\mu \ne 0$ are not components of an irreducible spherical
tensor. Instead, due 
to (\ref{pm1}) and (\ref{diag}) the relation $m'=m+\mu$ has to be
fulfilled:  
\begin{equation}
\{ M ^{\alpha,\alpha'}_{\mu, \mu'}\}^{l,l'}_{m,m'} (q;t) 
= \delta_{m+\mu,m'+\mu'} 
\{ M ^{\alpha,\alpha'}_{\mu, \mu'}\}^{l,l'}_{m,m'} (q;t)
\end{equation}

Without the memory matrix equation, (\ref{eq:mctanf}) would describe a system 
of coupled undamped harmonic "modes", where the modes are in this case 
correlation functions of tensorial densities.  
The physical origin of the memory matrix is the damping of these
oscillatory modes 
due to interaction between them including translation
rotation couplings, caused by the anharmonicities of the microscopic
interaction potentials. Of special importance will be the induction of 
a stress ${\cal L} \{j_1^T\}_{00}(q,t)/q$ by the force ${\cal L}
\{j_0^R\}_{21}(q,t)$ caused by the rotation of the molecules. This
mechanism is responsible for the existence of hydrodynamic
singularities in auto correlation functions of non hydrodynamic
fluctuations.    

\subsection{Molecular mode coupling theory}

Within the MMCT, the memory functions are
written as a sum of bare Markovian damping terms plus  mode coupling terms. 
The mode coupling terms have the form of self consistent statically 
renormalized one loop approximations. 
\begin{equation} \label{memory}
{\bf M}^{\alpha,\alpha'}_{\mu \mu'}(q;t-t') = i\,\mbox{\boldmath
$\nu$}_{\mu \mu'}^{\alpha \alpha'}(q) \delta(t-t') +
\frac{k_BT}{\Theta_{\alpha}} {{\bf m}}^{\alpha,\alpha'}_{\mu \mu'}(q;t-t')  
\end{equation}  

The derivation of the mode coupling approximations is analogous to the
one in \cite{schilling97}. For the memory functions with $\mu=\mu'=0$
the final result is identical to \cite{schilling97}. For
general $\mu,\mu'$, it can be written
\begin{eqnarray}
\label{mct}
& & \!\!\!\!\!\!\!\!\!\!\!\! 
\{m_{\mu,\mu'}^{\alpha, \alpha'}\}_{l,l'}^{m,m'}
(q;t)  \approx 
\frac{1}{2N} \left ( \frac{\rho_0}{4 \pi}\right ) ^2
{\sum _{\vec{q}_1 \vec{q}_2}}'
\sum_{m_1m_2}\sum_{l_1l_2}\;\sum_{l_1'l_2'} 
\times \nonumber \\
& & \times \, \{V^{\alpha \alpha '}_{\mu,\mu'}\}_{l,l',l_1,l_1',l_2, l_2'}^{m,m',m1,m2}(q,q_1,q_2)
\, S_{l_1 l_1'}^{m1}(q_1,t)\, S_{l_2l_2'}^{m2}(q_2,t) \; ,
\end{eqnarray}

with 

\begin{eqnarray}
\label{vertex}
& & \!\!\!\!\!\!\!\! 
\{V^{\alpha \alpha '}_{\mu,\mu'}\}_{l,l',l_1,l_1',l_2,
l_2'}^{m,m',m1,m2}(q,q_1,q_2)
:=  \nonumber \\
& & \{v^{\alpha}_{\mu}\}_{l,l_1, l_2}^{m, m_1, m_2} (q, q_1,q_2)\cdot
\{v^{\alpha'}_{\mu'}\}_{l',l_1', l_2'}^{m', m_1, m_2} (q, q_1,q_2)
^*\quad ,
\end{eqnarray}
\begin{eqnarray}
\label{vertex2}
& &  \!\!\!\!\!\! 
\{v^{\alpha}_{\mu}\}_{l,l_1, l_2}^{m, m_1, m_2} (q, q_1,q_2) :=
 \nonumber \\
& & \sum _{l_3}
\{u^{\alpha}_{\mu}\}_{l,l_3,l_2}^{m,m1,m2}(q,q_1,q_2)\,  
c_{l_3,l_1}^{m_1}(q_1)+ (-1)^m(1\longleftrightarrow 2)
\end{eqnarray}
where $c_{l,l'}^{m}(q)$ is the direct correlation function and 
\begin{displaymath} 
\{u^{\alpha}_{\mu}\}_{l,l_1,l_2}^{m,m1,m2}(q,q_1,q_2)
:=
i^{l_1+l_2-l}\left[\frac{(2l_1+1)(2l_2+1)}{(2l+1)} \right]^{\frac{1}{2}}
\frac{1}{2}\left[1+(-1)^{l_1+l_2+l}\right] \times
\end{displaymath}
\begin{eqnarray}
& \times & \sum_{m_1'm_2'}(-1)^{m_2'}d_{m_1'm_1}^{l_1}(\Theta_{q_1})
\,d_{m_2'm_2}^{l_2}(\Theta_{q_2})\,C(l_1l_2l;m_1'm_2'm) \nonumber \\
\label{mctend}
& \times & \left\{\begin{array}{l@{\quad;\quad}l}
q_1(\mu) \,C(l_1l_2l;000) & \alpha = T \\
\sqrt{l_1(l_1+1)}\,C(l_1l_2l;101) & \alpha = R \end{array} \right. \quad
\mbox{.}
\end{eqnarray}

Here the functions $q_1(\mu)$ are given by 
\begin{equation}
q_1(\mu) = q_1 \sqrt{\frac{4 \pi}{3}} \left (\mu \, Y_{1\mu}(\Theta_{q_1},\Phi_{q_1}) +
(1-| \mu |) \; 
Y_{10}(\Theta_{q_1},\Phi_{q_1}) \right ) 
\end{equation}
   
The functions $d_{m'm}^{l}(\Theta)$ are 
related to
Wigner's rotation matrices (we follow the notation of Gray and
Gubbins \cite{gray84}). For given Euler angles $\Phi,\Theta,\chi$ 
they are defined  as
\cite{gray84}  
\begin{equation}
\label{I2.3-2}
D_{mm'}^l(\Phi,\Theta,\chi)=e^{-im\Phi}\,d^l_{mm'}(\Theta)\,e^{-im'\chi}.
\end{equation}

$q_i,\Theta_{q_i},\Phi_{q_i}$ are the standard spherical coordinates
of $\vec{q}_i$ with respect to $\vec{q}$.
The prime at the first summation in Eq.(\ref{mct}) restricts
$\vec{q}_1,\vec{q}_2$ such, that $\vec{q}_1+\vec{q}_2 =\vec{q}$ in
order to fulfill momentum conservation. 
Eqs. (\ref{eq:mctanf}), (\ref{mct}), (\ref{vertex}) form a set of self
consistent equations for the generalized dynamic structure factors
$S_{ll'}^m(q,t)$ of linear molecules. They are slightly more general than
the equations in \cite{schilling97} by including the coupling to
transverse current fluctuations via a rotation - translation
coupling. It is to be expected that  this coupling will affect the
results for the glass
transition temperatures and the non ergodicity parameters studied in
\cite{schilling97,winkler00} only quantitatively but not
qualitatively \cite{theis00a}. 
The dynamics, instead, can be changed qualitatively in
certain wave vector  - ranges. Especially for small wave vectors the
hydrodynamic pole in the transverse current fluctuations  can
have large effects on the density relaxation spectrum. We will
demonstrate explicitly in the next chapter,  
that the coupling to transverse current fluctuations is necessary to 
reproduce the appearance of transverse sound modes in Brillouin
scattering spectrum of linear molecules, within the framework of MMCT.        

To calculate the light scattering
spectra it is most convenient to perform a Laplace Transform of Eq. 
(\ref{eq:mctanf}). With $LT(f(t))(z) = i \int_0^\infty e^{i z t} f(t)$,
with $Im(z) >0$, we obtain the following matrix equation
\begin{equation}
\label{eq:matrix1}
\left (
\begin{array}{cc}
z \underline{\underline{ I}} &
- \underline{\underline{\vec{q}}} \\
-\underline{\underline{\Gamma}}_{\vec{j} \rho} &
z \underline{\underline{ I}} + \underline{\underline{M}}_{\vec{j} \vec{j}}
\end{array}
\right ) \left (
\begin{array}{cc}
\underline{\underline{\phi}}_{\rho \rho}(z) &
\underline{\underline{\phi}}_{\rho \vec{j} }(z)\\
\underline{\underline{\phi}}_{\vec{j} \rho}(z) &
\underline{\underline{\phi}}_{\vec{j} \vec{j}}(z) 
\end{array} 
\right ) = - \left (
\begin{array}{cc}
\underline{\underline{\phi}}_{\rho \rho}^0 &
\underline{\underline{\phi}}_{\rho \vec{j}}^0 
\\
\underline{\underline{\phi}}_{\vec{j} \rho}^0&
\underline{\underline{\phi}}_{\vec{j} \vec{j}}^0
\end{array} \right )
\end{equation}

Here we have chosen a simplified notation. In the q - frame the first
matrix in Eq. (\ref{eq:matrix1}) would be explicitly: 
\begin{equation}
\label{eq:matrix2}
\left (
\begin{array}{cc}
z \underline{\underline{ I}} &
- \underline{\underline{\vec{q}}} \\
-\underline{\underline{\Gamma}}_{\vec{j} \rho} &
z \underline{\underline{ I}} + \underline{\underline{M}}_{\vec{j} \vec{j}}
\end{array}
\right ) =
\left (
\begin{array}{cc}
z \, \delta_{ll'}\, \delta_{mm'} &
- \delta_{ll'} \, \delta_{mm'} \, (-1)^\mu q_{-\mu}^\alpha(l) \\
-\delta_{\mu,0}\, \frac{k_B T}{\Theta_\alpha} q_0^\alpha \,
\delta_{mm'} \,
({\bf S}^{-1}) (q)  &
z \, \delta_{ll'}\, \delta_{mm'} + {\bf M}_{\mu^,\mu'}^{\alpha
\alpha'}(q,z)  
\end{array}
\right )
\end{equation}
The matrix of static correlators on the right hand side of
Eq. (\ref{eq:matrix1}) is
\begin{equation}
\left( \begin{array}{cc}
\underline{\underline{\phi}}_{\rho \rho}^0 &
\underline{\underline{\phi}}_{\rho \vec{j}}^0 
\\
\underline{\underline{\phi}}_{\vec{j} \rho}^0&
\underline{\underline{\phi}}_{\vec{j} \vec{j}}^0
\end{array} \right )
=
\left( \begin{array}{cc}
\delta_{mm'} S_{ll'}^m(q) & 0
\\
0&
\delta_{\mu \mu'} \delta_{\alpha \alpha'} \delta_{mm'} \delta_{ll'}
\frac{k_B T}{\Theta_\alpha}
\end{array} \right )
\end{equation}

It is obvious from equations (\ref{eq:matrix1}) and (\ref{eq:matrix2}) that
the sum rules (\ref{ivvtot}) and (\ref{ivhtot}) are automatically
fulfilled, if the approximations for the memoryfunctions obey the very
weak requirement, that $\lim_{z \to \infty + i
\epsilon}\frac{\{M_{\mu\mu'}^{\alpha\alpha'}\}_{ll'}^{mm'}}{z} = 0 $, 
where $\epsilon$ is an arbitrary  positive number. Since memory
functions are regular for $t=0$ in most
physical cases, they even vanish as $1/z$ for $z \to \infty$. 
 This property is especially fulfilled for the selfconsistent MMCT and
also for the primitive theory used in section (\ref{sec:glasd}). 

\section{Glassy dynamics and hydrodynamic modes}
\label{sec:glasd}

Light scattering usually measures at small wave vectors
$q$. Therefore the correct treatment of hydrodynamic modes becomes
crucial for  explaining light scattering experiments. To obtain the
generalized density correlation $S_{ll'}^m(q,t)$ for small
wave-vectors, it would be necessary to solve the self consistent set of
equations (\ref{eq:mctanf},\ref{mct},\ref{vertex}) for {\em all}
wave-vectors and {\em all} $l,m$, since all degrees of freedom are
coupled via mode coupling integrals. To study the light scattering
problem, we first want to restrict the discussion to the fluctuations, 
which are most relevant for the understanding of light scattering
experiments. These are the density fluctuation $\rho_{lm}(q,t)$ for
$(l,m)=(0,0)$ and $2,m$ their
respective current fluctuations
$\{j_0^T\}_{00}(q,t),\{j_0^R\}_{2m}(q,t)$, 
which are
directly measured in light scattering experiments (i.e. $l=0,2$ and
$m=0,1,2$) and the current fluctuations $\{j_\mu^T\}_{00}(q,t)$, which
are slow,  
since the total  currents $\{j^T_\mu\}_{00}(q=0,t)$ are conserved. Here
we have also neglected the current fluctuations
$\{j^T_0\}_{2m}(q,t)$. Their contribution will be of higher order in
$q$, as can be seen by comparing the translational and
rotational current contribution in the first line of
(\ref{eq:mctanf}) together with (\ref{eq:qmu}).
Additional simplification occur in
the limit $q \to 0$, due to relation (\ref{diag}) for correlation
functions of spherical tensors in a rotationally invariant system.   
\begin{enumerate}
\item The static correlation function is $S_{ll'}^m(q=0) =
  \delta_{ll'}\; \; 
S_l \equiv S_{ll}^m(q=0)$, independent of m.   
\item   The component of the matrix
$\underline{\underline{\Gamma}}_{\vec{j} \rho }$ of
Eq. (\ref{eq:matrix1}) with $l=l'= m= \mu=0$ and $\alpha=T$ reduces to  
 $\frac{k_B T}{m S_{00}(q)}  q  =  c_\parallel^2 q $. 
Here $c_{\parallel}$ is the longitudinal isothermal sound velocity in the
liquid. 
\item
The  component of the matrix
$\underline{\underline{\Gamma}}_{\vec{j} \rho }$ of
Eq. (\ref{eq:matrix1}) with $l=l'=2$,  $\mu=0$ and $\alpha=R$ reduces to 
 $\frac{k_B T}{\Theta S_2(q)}  \sqrt{6}  =
\frac{\omega_R^2}{\sqrt{6}} $,  where $\omega_R$ is a classical
frequency, related to the rotation of the quadrupoles.   
\item
The memory matrix $\{M^{TT}_{00}\}_{00}^{00}(i 0)$ has to reduce to $i \eta_l
q^2$, where $\eta_l$ is the longitudinal viscosity of the liquid. The
transversal memory function is for $q \to 0, z \to i 0$ given by
$\{M^{TT}_{11}\}_{00}^{00}(i0) = i q^2 \eta_S$, where $\eta_S$ is the
shear viscosity. The $q^2$ dependence of the translational memory functions
$\{m_{\mu\mu'}^{TT}\}_{00}^{00}$  are due to
momentum conservation. The parameter
$\{K_{\mu\mu'}^{TT}\}_{00}^{00}$ and
$\{\tau_{\mu\mu'}^{TT}\}_{00}^{00}$ in (\ref{maxwell}) are 
identical to the longitudinal modulus $K_l$ and the 
longitudinal  $\alpha$ -
relaxation time $\tau_l$ for $\mu=\mu'=0$ and the shear modulus
$G_S$ and the transverse $\alpha$ - relaxation time
$\tau_S$ for $\mu=\mu'=1$, respectively.  With this choice the
Maxwell relations $\eta_l =  \eta_l^0 + K_l \tau_B$ and $\eta_S=
\eta_S^0 + G_S \tau_S$, where $\eta_l^0$, $\eta_S^0$ are the
contribution from the Markovian part of the memory matrix, 
are fulfilled in the liquid. 
\end{enumerate}

Due to the local nature of the
cage effect, which is responsible for the slowing down of structural
relaxations, it is strictly speaking not possible to study
self-consistently the
hydrodynamic limit, without the knowledge of relaxations on local
length scales $r \propto r_0$, where $r_0$ is on the scale of intermolecular
distances.  But since the memory functions ${\bf
  m}_{\mu,\mu'}^{\alpha,\alpha'}(q,t)$  do not contain any
hydrodynamic pole by construction, they are non trivial only due to
the glass transition dynamics,  which in turn is independent of the
hydrodynamic fluctuations at short wavelength. It is e.g. theoretically
understood \cite{leshouches} and verified in
simulations \cite{gleim},
that systems with qualitatively different hydrodynamic behavior exhibit
the same glassy dynamics. To obtain the qualitative behavior of
Eqs. (\ref{eq:mctanf},\ref{mct},\ref{vertex}), it is therefore
sufficient to replace the mode coupling part of the memory function
matrix by its leading wave-vector
behavior multiplied with a function, which is
able to describe glassy dynamics. Although the memory functions are
free of hydrodynamic singularities, they exhibit the full frequency
dependence of glassy dynamics (``fast'' and ``slow'' $\beta$ -
relaxations, $\alpha$ - relaxation plus additional complications as
e.g. contributions from Bose peak phenomena
\cite{goetze00,theenhaus00}). For reproducing all details of light
scattering spectra, which are directly related to glassy dynamics, two
approaches are possible. 
Either an ansatz has to be found, which is compatible with all the
mentioned phenomena, or  the full set of microscopic equations,
derived in sec. \ref{sec:mmct}, had to be solved numerically, to
account at least for $\beta$ - and $\alpha$ - relaxation, (see
e.g. \cite{goetze92}).  But before
this very difficult problem can be treated, it is necessary to
demonstrate, that the structure of the equations derived in sec
\ref{sec:mmct} can account  for all the hydrodynamic poles and
their interplay with the most basic phenomena of structural
relaxations i.e the $\alpha$ - relaxation. To achieve that, it is
sufficient to use an $\alpha$ - relaxation ansatz for the
memory function. To make the analysis as simple and explicit as
possible, we choose simple exponentials (Maxwell theory) for the non
vanishing memory functions, multiplied with their leading wave vector
dependence, $q^n$, $n \in \{0,1,2\}$. 

\begin{equation} \label{maxwell}
\{m_{\mu,\mu'}^{\alpha,\alpha'}\}_{l,l'}^{m,m'}(q,z) = -
\frac{q ^n \{K_{\mu,\mu'}^{\alpha,\alpha'}\}_{l,l'}^{m,m'}(q) \; \; 
\{\tau_{\mu,\mu'}^{\alpha,\alpha'} \}_{l,l'}^{m,m'}}{z \;
\{\tau_{\mu,\mu'}^{\alpha,\alpha'} \}_{l,l'}^{m,m'}  + i}       
\end{equation}
The small wave vector behavior of the memory functions (\ref{maxwell})
can be derived from (\ref{diag}) and (\ref{pm1}) and the conservation
laws for total momentum in every spatial direction 
for $l=0$ or $l'=0$. The  relaxation times
$\{\tau_{\mu,\mu'}^{\alpha,\alpha'}\}_{l,l'}^{m,m'}$ are taken at
$q=0$. The projection operator formalism guarantees, that this value is
nonzero, since the memory functions do not contain hydrodynamic
poles. The Markovian part of the memory functions is in the following
neglected, if it would vanish at wave vector $q=0$. There are two
severe  
consequences of this approximation together with the $\alpha$ - 
relaxation ansatzes for the memory functions. First the sound poles in the
glass do not show any damping, instead of a damping proportional to
$q^2$. Second, a fit with the Maxwell ansatz or any other ansatz, which only
describes the $\alpha$ - relaxation, 
would lead to an artificial time scale separation for the 
$\alpha$ - relaxation times  of the different memory functions (see
e.g. the discussion in \cite{dreyfus99}). The source of both errors,
is the neglect of "fast" $\beta$ - relaxation phenomena, which also
contribute to the memory functions.

The task is now to show, that the predicted spectra are consistent
with sum rules and the qualitative behavior of
light scattering experiments, which already follow very generally from a purely
generalized hydrodynamic analysis \cite{franosch00} and that the
interplay of rotational and translational motions lead to the
qualitatively correct renormalizations of the hydrodynamic poles, when
entering the glassy regime.

If we order our basic variables in the form
$\rho_{00},\rho_{2m},\{j_0^R\}_{2m},\{j_1^T\}_{00}$,
the
frequency independent matrices appearing in Eq. (\ref{eq:matrix1}) are 
\begin{equation}
\underline{\underline{\vec{q}}} = 
\left(\begin{array}{ccccc} 
q&&&&0\\
&\sqrt{6}&&&0\\
&&\sqrt{6}&&0\\
&&&\sqrt{6}&0
\end{array} \right)
\quad \quad  \underline{\underline{\Gamma_{\vec{j}\rho}}} = 
\left(\begin{array}{cccc} 
c_\parallel^2 q&&&\\
&\omega_R^2/\sqrt{6}&&\\
&&\omega_R^2/\sqrt{6}&\\
&&&\omega_R^2/\sqrt{6}\\
0&0&0&0
\end{array} \right)
\label{eq:rot}
\end{equation}
The slow part of the  memory matrix is given by 
\begin{equation}
 z \underline{\underline{I}} + \underline{\underline{m}}_{\vec{j} \vec{j}} = 
\left (
\begin{array}{ccccc}
z-\frac{q^2 K_l \tau_l}{z \tau_l + i} & -\frac{q K_{lR} \; \tau_{lR} }{ z\;
  \tau_{lR} +i} &&& \\
-\frac{q K_{lR} \; \tau_{lR} }{ z
  \tau_{lR} +i}  & z - \frac{K_R \;\tau_R}{z \;\tau_R+i } &&& \\
&& z  - \frac{K_R\; \tau_R}{z \tau_R+i }& & -\frac{q K_{SR}\;
  \tau_{SR}}{ z \;
  \tau_{SR}+i}  \\
&&& z - \frac{K_R \; \tau_R}{z \; \tau_R+i }& \\
&& -\frac{q K_{SR} \; \tau_{SR}}{ z \tau_{SR}+i} && z-\frac{q^2 G_S
  \tau_S}{z \tau_S + i}  
\end{array}
\right )
\end{equation}

The leading wave-vector dependence for $q \to 0$ of the matrix elements
is derived from
conservation laws and Eq. (\ref{diag}). All the matrix elements, which
are left empty are exactly zero due to $m$,$ m'$ selection rules (see
Eq. (\ref{diag})). 
For the appearance of
transversal sound modes in the light scattering spectra and the
explanation of the Rytov Dip it is crucial that $K_{SR} \ne 0$, which
quantifies the memory matrix-element between the transversal current
for $l=0$ and the rotational current $j_{2m}^R$ for $m=1$. The four
parameters $K_l$, $K_R$, $K_{lR}$ and $ K_{SR}$ have to be such 
that the memory matrix remains positive definite for all frequencies. 
For $z=0$ the relation
\begin{equation}
K_l K_R \;\tau_l \tau_{R} -  K_{lR}^2 \;\tau_{lR}^2 > 0 \;\;\; \mbox{and}
\;\;\;  K_R G_S \;\tau_S \tau_R - K_{SR}^2 \;\tau_{SR}^2> 0
\end{equation}
follow and the diagonal elements $K_l$,$K_R$ and $G_S$ have to be positive.
We also note that the exact relations 
\begin{eqnarray}
G_s K_R \ge K_{SR}^2 && \label{cs}\\
K_l K_R \ge K_{lR}^2&&\label{cl}
\end{eqnarray}
can be derived from the Cauchy relations 
\begin{eqnarray}
{\{m_{11}^{TT}\}_{00}^{00}}''(\omega)
{\{m_{00}^{RR}\}_{22}^{00}}''(\omega) \ge
{\{m_{10}^{TR}\}_{02}^{00}}''(\omega)^2&& \label{cauchys}\\
{\{m_{00}^{TT}\}_{00}^{00}}''(\omega)
{\{m_{00}^{RR}\}_{22}^{00}}''(\omega) \ge
{\{m_{00}^{TR}\}_{02}^{00}}''(\omega)^2&& \label{cauchyl}
\end{eqnarray} 
by considering the low and high frequency limits of
Eqs. (\ref{cauchys}) and (\ref{cauchyl}).  

\subsection{Hydrodynamic poles}
Before we discuss the numerical solution of Eq. (\ref{eq:matrix1}), it is
useful to investigate its hydrodynamic poles.  The only conserved 
quantities   are
(besides the total energy)  the center of mass density and the total
momentum in every spatial direction. They are the cause of the
hydrodynamic poles (i.e. poles which show dispersion laws $z \propto
q^n$, with $n=1,2$) in the respective auto correlation functions
$(\Phi_{\rho \rho})_{00}^0$ and
$\{(\Phi_{\vec{j}\vec{j}})_{\mu\mu}^{TT}\}_{00}^{00}\}$. But due to
the dynamic coupling of   the rotational degrees of freedom and the
translational degrees of freedom, which appear naturally in the memory
matrix, also the correlation functions of non
hydrodynamic variables do exhibit hydrodynamic poles. To study this
phenomena, we 
invert the matrix in Eq. (\ref{eq:matrix1}), use the ansatz $z=p
q^n $, $n \in \{ 0,1,2\} $ and expand the denominator in powers of
$q$. For $n=0$ this gives poles of non--hydrodynamic nature (
rotational modes with a frequency $p$), for $n=1$ propagating modes
(transverse or longitudinal phonon modes) are described with $p$ being
the sound velocity and for $n=2$ a diffusive mode with a transport
coefficient (in our case generalized
viscosity) $-p/i$ is obtained.  

Let's first study the transversal current fluctuations. In a {\em simple}
liquid the transversal current correlator $\{(\Phi_{\vec{j}
\vec{j}})_{11}^{TT}\}_{00}^{00}(z)$ exhibits a viscous pole 
at $z = -i G_S \tau_S q^2$. This is also the case for the liquids of
linear molecules, studied in this paper. 
But in addition, due to the
translation rotation coupling, also the auto correlation functions of densities
$\rho_{lm}(q,t)$ or currents $\{j_0^T\}_{lm} $ with $l=2$, $m=1$
do exhibit the transversal hydrodynamic poles. The reason for that is, 
that the tensor $\{j_1^T\}_{00} $ has irreducible spherical
components with $l=2$,$m=1$, and thus is able to couple dynamically
to the
specified tensors via the memory functions with $m=1$. There is also a
non-hydrodynamic singularity related to the 
rotational motion of the molecules. 
If we restrict the correlators to
their poles (neglecting the glassy dynamics), the poles of the $
(\Phi_{\rho  \rho})_{22}^1(z),\; 
\{(\Phi_{\vec{j}  \vec{j}})_{00}^{RR}\}_{22}^{11}(z), \;\{(\Phi_{\vec{j}
\vec{j}})_{11}^{TT}\}_{00}^{00}(z)$ components are given by:
\begin{equation}
\left ( z^2 + i (K_R \tau_R  + \nu_R) z -\omega_R^2 \right )
\; \left ( z + i G_S\, \tau_S q^2 \right ) = 0
\end{equation}
The two poles couple into the dynamic correlators with different
amplitudes. The coupling of the second pole into the $l=2$, $m=1$ rotational
component causes the Rytov dip. As an example we therefore give the
the term in lowest order of $q$ 
of the strength of the transverse sound mode coupling into the
$(\Phi_{\rho  \rho})_{22}^1(z)$ component. In the vicinity of the
Rytov dip this component can be expressed as:
\begin{equation}
(\Phi_{\rho  \rho})_{22}^1(z) =
\frac{- K_{SR} \tau_{SR} q^2}{\omega_R^2} \left (
\frac{-1}{z + i G_S\, \tau_S q^2} \right ) 
\end{equation}
Therefore the strength of the Rytov dip is proportional to $q^2$ times
the matrix element which couples the transverse current to the $m=1$
rotational motion and vanishes if the rotational frequency $\omega_R$
goes to infinity.

As  soon as $z\tau_S \gg 1$, the
diffusive pole will turn into a propagating transversal
sound mode. In simple liquids this pole will be at $z = \pm \sqrt{G_S}
q$. Whereas in {\em molecular} liquids the
transversal sound velocity is renormalized by contributions of the
rotational degrees of freedom.  The pole structure of the specified
correlators 
$(\Phi_{\rho  \rho})_{22}^1(z),\; 
\{(\Phi_{\vec{j}  \vec{j}})_{00}^{RR}\}_{22}^{11}(z), \;\{(\Phi_{\vec{j}
\vec{j}})_{11}^{TT}\}_{00}^{00}(z)$ is in the supercooled regime 
given by:
\begin{equation}
\left ( z^2  + i \nu_R z - \omega_R^2 - K_R \right )
\left ( z^2 - (G_S - \frac{K_{SR}^2}{(K_R + \omega_R^2)}) q^2 \right )
= 0
\end{equation} 
I.e. the transversal sound pole is given by
\begin{equation}
\label{eq:cperp}
z= \pm\sqrt{G_s  - \frac{K_{SR}^2}{(K_R + \omega_R^2)}} \; q \;\; := \pm 
c_\perp q 
\end{equation}
The transverse sound velocity is shifted to smaller frequencies
compared to what is expected in a 
simple liquid. This trend was already noted in \cite{dreyfus99}. Note, 
that due to the positivity of $\omega_R^2$ and the exact relation
(\ref{cs}), the transversal sound velocity is always well defined. 
I.e. by treating the rotation translation coupling explicitly, we are 
able to describe the contribution of the rotational motion to the
transversal sound velocity $c_\perp$. The
microscopic rotational translational coupling is the cause of
the appearance of hydrodynamic poles in correlation functions of non
hydrodynamic (i.e. for $q \to 0$ non conserved) variables and of a
renormalization of the transversal sound velocity. 
Analogous behavior is found for center of mass and longitudinal
current fluctuations. The amplitude of the transverse sound pole in
lowest order of $q$ which occurs in the 
$(\Phi_{\rho  \rho})_{22}^1(z)$
component can be derived. It is in the vicinity of the transverse
phonon mode given by:
\begin{equation}
(\Phi_{\rho  \rho})_{22}^1(z)
= \left ( 
\frac{ K_R^2 ( G_S ( K_R + \omega_R^2))}
{(K_R + \omega_R^2)(K_SR^2(\omega_R^2 - K_R) + G_S K_R (
K_R+\omega_R^2))}
\right ) \frac{-1}{z \pm \sqrt{G_s  - \frac{K_{SR}^2}{(K_R +
\omega_R^2)}} \; q} 
\end{equation}
Therefore the transverse phonon mode can only be observed as long as
the rotation couples via $K_R$ to the structural relaxation.

In the fluid ($\tau_l c_{\parallel} q \ll 1$) the longitudinal
components with $l=0, m=0$ and the rotational components with $l=2,
m=0 $ contain two types of modes . First the $l=2$ rotational mode and
second the longitudinal phonon mode. The poles of $
(\Phi_{\rho  \rho})_{00}^0(z),\; (\Phi_{\rho  \rho})_{22}^0(z),\;
\{(\Phi_{\vec{j}  \vec{j}})_{00}^{TT}\}_{00}^{00}(z), \;\{(\Phi_{\vec{j}
\vec{j}})_{00}^{RR}\}_{22}^{00}(z)$ are given by:
\begin{equation}
\label{eq:longphon}
\left ( z^2 +
i (K_R \tau_R + \nu_R) z - \omega_R^2 \right )  
\left ( z^2 - c_{\parallel}^2 q^2 + i K_l \tau_l z q^2 \right )
= 0
\end{equation}
The first pole gives  the
damped rotation of the molecule, whereas the second term describes 
the usual longitudinal sound modes in the liquid at $z_\pm = \pm
c_{\parallel} q$ with the damping $K_l \tau_l q^2/2$. This expression
is valid as long as $ K_l \tau_l q^2 \ll \tau_l c_{\parallel} q \ll 1 $. 
The amplitude of the longitudinal phonon mode in $(\Phi_{\rho
\rho})_{22}^0(z)$ is proportional to $q^2$, where it is of order $q^0$
in $(\Phi_{\rho  \rho})_{00}^0(z)$. Its contribution to the sound pole
in the $I^{VV}$ - spectrum in the liquid can therefore be neglected.  


In the solid ($\tau_l c_{\parallel} q \gg 1$), the sound pole will be
shifted to higher frequencies and, as  an artifact of the
Maxwell theory,  the damping  vanishes. An inclusion of $\beta$
relaxation phenomena  will cure this 
unphysical behavior.  We obtain as poles of the $(\Phi_{\rho  \rho})_{00}^0(z),\; (\Phi_{\rho  \rho})_{22}^0(z),\;
\{(\Phi_{\vec{j}  \vec{j}})_{00}^{TT}\}_{00}^{00}(z), \;\{(\Phi_{\vec{j}
\vec{j}})_{00}^{RR}\}_{22}^{00}(z)$ components as:
\begin{equation}
\left (
z^2 + i \nu_R z -\omega_R^2 - K_R \right )
\left ( z^2 - (c_{\parallel}^2 + K_l - \frac{K_{lR}^2}{(K_R +
\omega_R^2)})q^2
\right )
= 0
\end{equation}
i.e the longitudinal sound
velocity $c_{\infty} $ is, as the transversal sound velocity, modified
by rotational degrees of
freedom. 
\begin{equation}
\label{eq:longsol}
c_{\infty}^2 = c_{\parallel}^2 + K_l - \frac{K_{lR}^2}{(K_R +
\omega_R^2)}
\end{equation}
Due to the positivity of $\omega_R^2$ and the exact relation
(\ref{cl}) the sound velocity is always shifted to higher values in
the glass, but the shift is reduced compared to what would be expected 
in a simple liquid. 
In analogy to the transverse mode we can give the low $q$ expansion
for the amplitude of the longitudinal sound pole in the $(\Phi_{\rho
\rho})_{22}^0(z)$. In the vicinity of the longitudinal phonon
frequency this component is given by:
\begin{equation}
(\Phi_{\rho \rho})_{22}^0(z) = 
\left (
\frac{K_R}{K_R+\omega_R^2}
\right )
\frac{-1}{z \mp \sqrt{ c_{\parallel}^2 + K_l - \frac{K_{lR}^2}{(K_R +
\omega_R^2)}}}
\end{equation}
Therefore a longitudinal phonon in the $(\Phi_{\rho
\rho})_{22}^0(z)$ component is always observable in the
supercooled liquid as long as the rotational motion couples via the
matrix element $K_R$ to the structural relaxation.

\subsection{A solution of the equation of motion}
\label{sec:sec4fig}

We have now solved the equation of motion for some chosen but fixed
parameters. Close to the glass transition it is only the scale
of the $\alpha$ - 
relaxation time, which is changing considerably. If the time
temperature superposition principle (TTS) is fulfilled exactly all
relaxation times $\{\tau_{\mu \mu'}^{\alpha \alpha'}\}_{ll'}^{mm'}(T)$ can 
be written  as $\{\hat{\tau}_{\mu \mu'}^{\alpha \alpha'}\}_{ll'}^{mm'} 
f(T)$.  The function $f(T)$ is the same for all relaxation times and
is a quickly changing function of temperature. In the idealized mode
coupling theory it would be $(|T-T_c|/T_c)^{-\gamma}$, often used fit
formulas are the Vogel Fulcher function $f(T) = A \exp(-B/(T-T_0))$ or
as a special case the Arrhenius law ($T_0=0$). The prefactors
$\{\hat{\tau}_{\mu \mu'}^{\alpha \alpha'}\}_{ll'}^{mm'}$ are constant, 
but in general different for different combinations of
$\mu,\mu',\alpha,\alpha',l,l',m ,m'$.  Very often the TTS is
violated at lower temperatures. In this case also the
$\{\hat{\tau}_{\mu \mu'}^{\alpha \alpha'}\}_{ll'}^{mm'}$ vary slowly
with temperature.  

Since we are in this paper only interested in qualitative aspects of
the solution, all $\alpha$ - relaxation times were taken  equal 
i.e. $\{\hat{\tau}_{\mu \mu'}^{\alpha \alpha'}\}_{ll'}^{mm'} =1$. The
function $f(T)$ is called $\tau$ in the following. 
In this way we demonstrate how
we can explain the complete viscosity range from the highly viscous
liquid towards the glass by just varying a single parameter, $\tau$.

\begin{figure}
\unitlength1cm
\epsfxsize=10cm
\begin{picture}(9,8)
\put(-0.5,-1){\rotate[r]{\epsffile{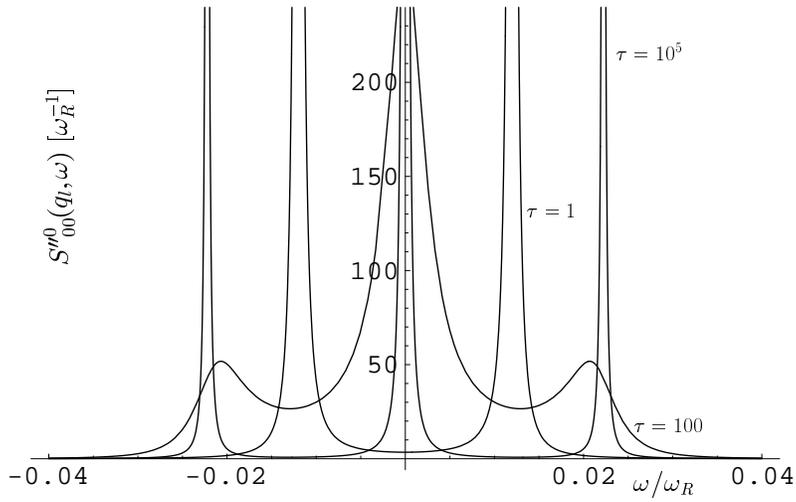}}}
\end{picture}
\caption{Solution of Eq. (\protect\ref{eq:matrix1}) for the
dynamical structure factor
${S''}_{00}^0(q,\omega)$ 
The parameters where set as for Fig. \protect\ref{fig:dep}}
\label{fig:1}
\end{figure}

\begin{figure}
\unitlength1cm
\epsfxsize=10cm
\begin{picture}(9,8)
\put(-0.5,-1){\rotate[r]{\epsffile{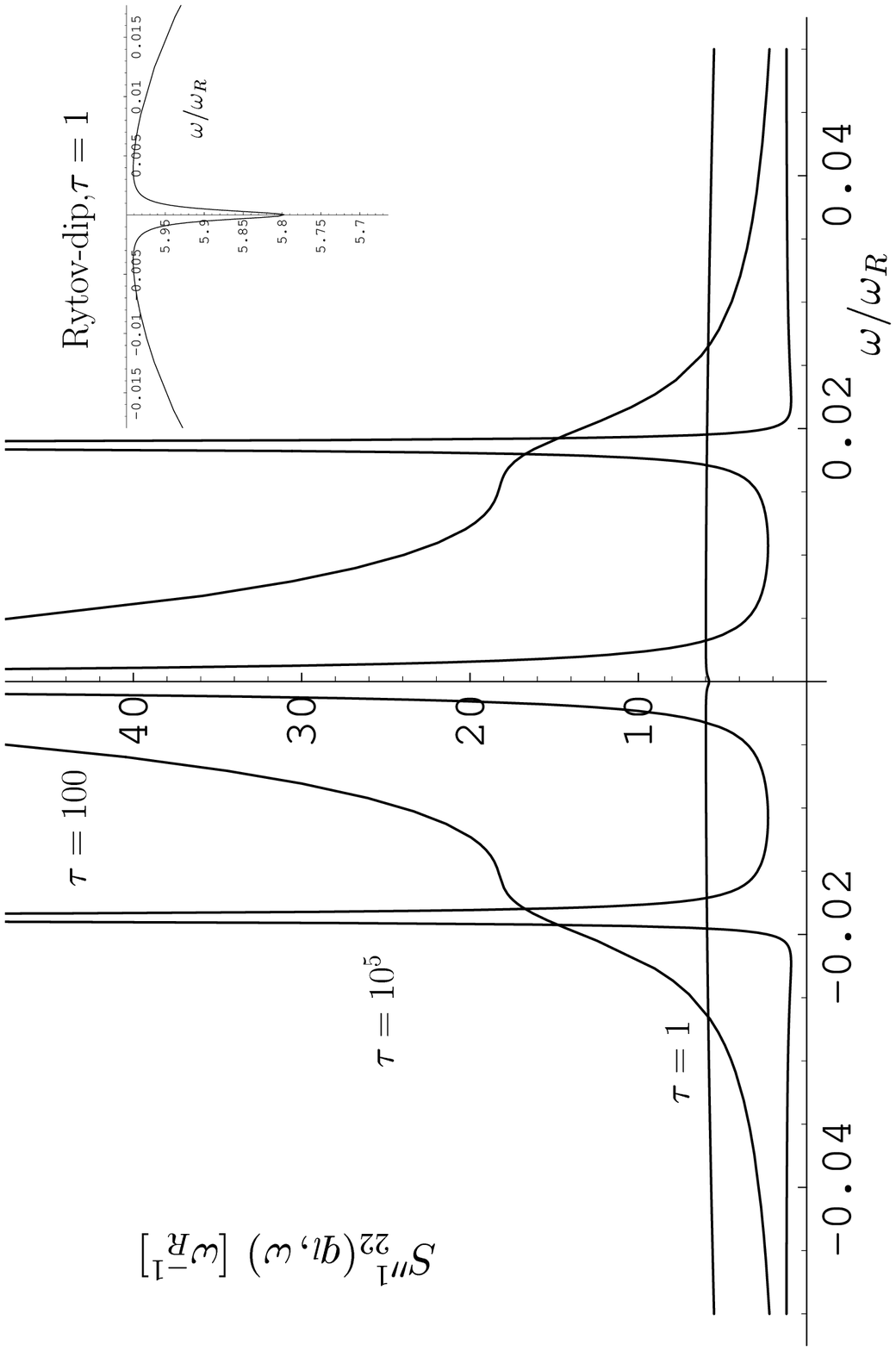}}}
\end{picture}
\caption{The spectrum ${S''}_{22}^1(q,\omega)$ for the same parameters
as in Fig. \protect\ref{fig:1}.  The inset shows the value around zero
frequency for $\alpha = 
1$ to show the Rytov dip.
}
\label{fig:2}
\end{figure}

\begin{figure}
\unitlength1cm
\epsfxsize=10cm
\begin{picture}(9,8)
\put(-0.5,-1){\rotate[r]{\epsffile{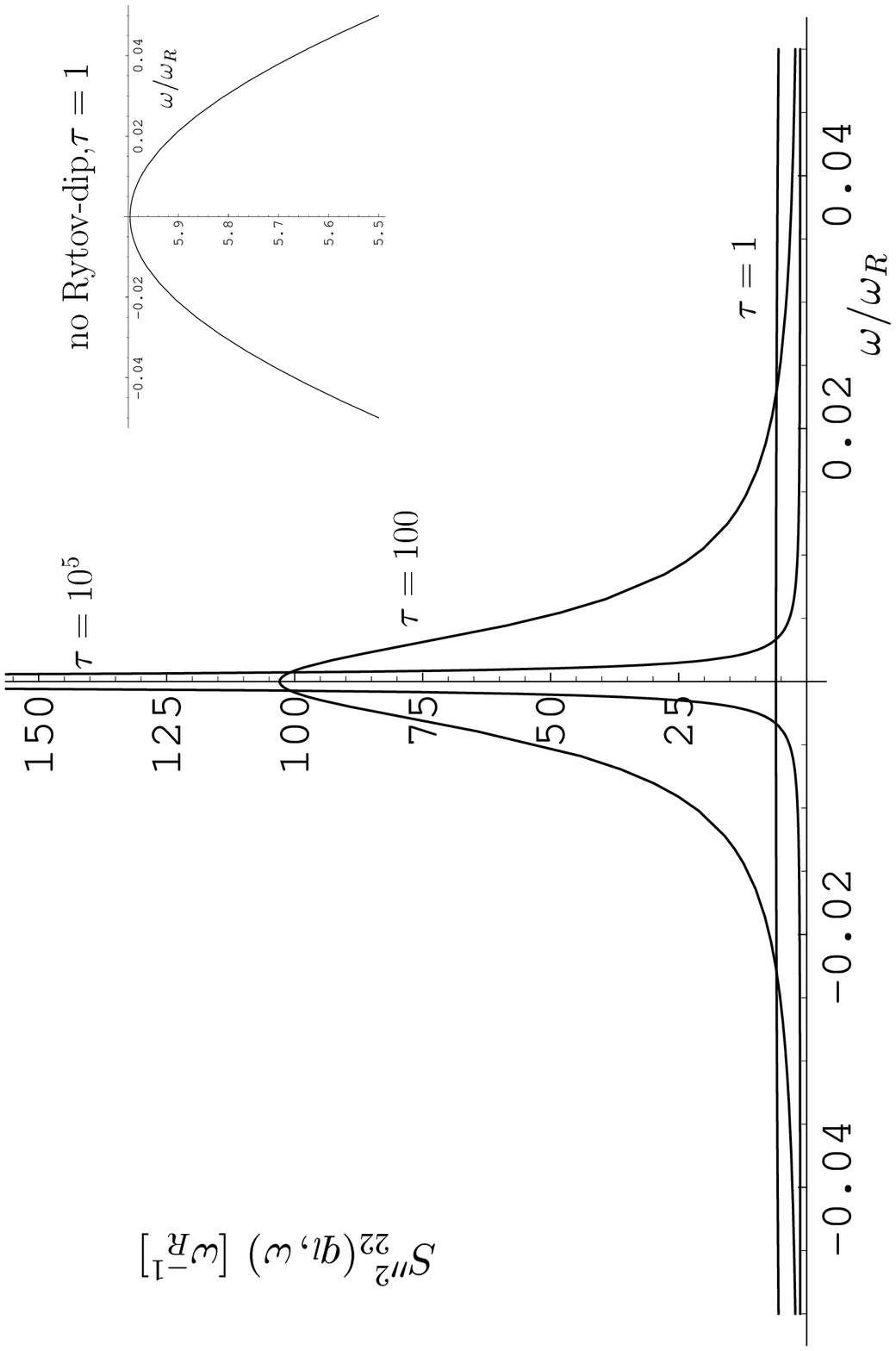}}}
\end{picture}
\caption{The spectrum ${S''}_{22}^2(q,\omega)$ for the same parameters
as in Fig.  \protect\ref{fig:1}. The inset shows the value around zero
frequency for $\alpha = 
1$ to show the absence of the Rytov dip.
}
\label{fig:3}
\end{figure}

\begin{figure}
\unitlength1cm
\epsfxsize=10cm
\begin{picture}(9,8)
\put(-0.5,-1){\rotate[r]{\epsffile{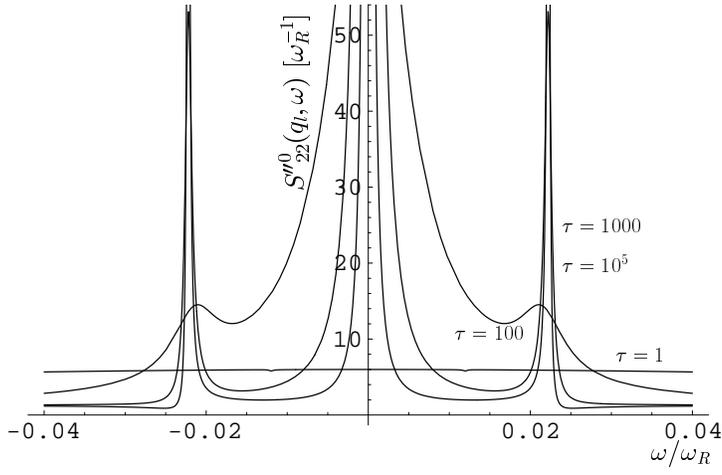}}}
\end{picture}
\caption{The
spectrum ${S''}_{22}^0(q,\omega)$ for the same parameters as in
Fig. \protect\ref{fig:1}. 
The longitudinal phonon couples into this component and
becomes observable when the light scattering experiment starts to
probe a solid. 
}
\label{fig:220}
\end{figure}

\begin{figure}
\unitlength1cm
\epsfxsize=10cm
\begin{picture}(9,8)
\put(-0.5,-1){\rotate[r]{\epsffile{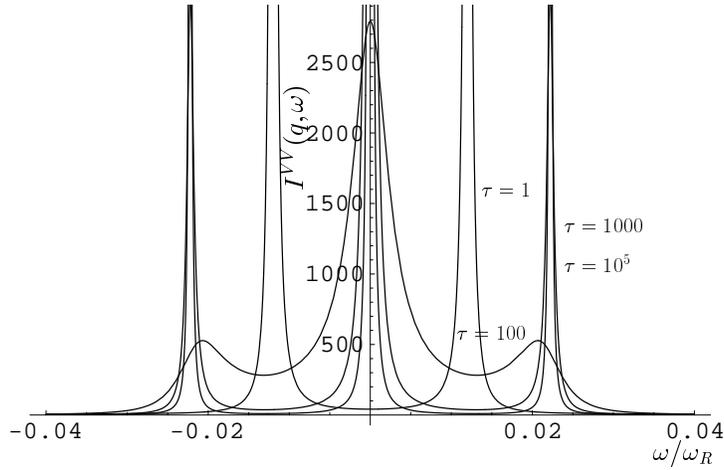}}}
\end{picture}
\caption{The VV spectrum for the same parameters as in
Fig. \protect\ref{fig:1}.  
The square of the isotropic part of the polarizability was assumed to
be 10 times as 
big as the isotropic one $a^2=10g^2$.
$I^{VV}(q,\omega) \sim a^2 {S''}_{00}^0(q,\omega) + \frac{4 \pi}{15} g^2
\left ( {S''}_{22}^2(q,\omega) + \frac{1}{3}  {S''}_{22}^0(q,\omega)
\right )$.
}
\label{fig:VV}
\end{figure}

In the figures \ref{fig:dep} - \ref{fig:VVlog} we have chosen the
following parameters: The frequency scale for the rotational frequency
was set to unity $\omega^R=1$. In these units the other parameters
where chosen as $c_{\parallel}=0.6$, $G_S=1$, $\nu_R=5$, $K_l=1$,
$K_R=1$, $K_{lR}=K_{SR}=1/2$. The external momentum $q$ selected by the
scattering experiment is set to $q=0.02$. For $z \tau \ll 1$ 
the scattering experiment probes an amorphous solid whereas for  
$z \tau \gg 1$ it probes a liquid.


In Fig. \ref{fig:1} we have plotted from a solution of
Eq. (\ref{eq:matrix1}) the imaginary part of the center of mass
correlator $S_{00}^0(q,z)$ which 
according to Eq. (\ref{eq:vv})
forms the main part of the polarized light scattering intensity
$I^{VV}(\omega)$. For $\tau = 1 \ll (c_{\parallel} q)^{-1}$ the
Brillouin line, caused by a  longitudinal sound wave, is at
$c_{\parallel} q = 0.012$. The damping is proportional to $q^2$ (see
Eq. (\ref{eq:longphon})).
For 
$\tau = 100 \approx  (c_{\parallel} q)^{-1} $ a broad central peak occurs
together with a strong damping of the phonon modes. This is a well
known phenomena which was already explained by Mountain
\cite{mountain66} for the
Brillouin spectrum. 
The same mechanism applies here. 
If we calculate the hydrodynamic sound pole with the condition $\omega \tau
=1$ the equation for the sound pole is 
\begin{equation}
\omega^2 + i \omega \nu q^2  + i \frac{K_l}{2} q^2 - (c^2_\parallel +
\frac{K_l}{2}) q^2 = 0
\end{equation}
This equation can be solved with the ansatz $\omega = c_\infty q + i
\Gamma q + O(q^2)$ i.e. the damping is of order $q \gg q^2$ instead of
order $q^2$.  This effect is even stronger in real glass-formers, where 
the $\alpha$ - relaxation is better described by a stretched
exponential behavior instead of a single exponential. We can estimate, 
that for frequencies in the high
frequency wing of the $\alpha$ - relaxation, where the von Schweidler
law applies for the memory-functions ($m(z) \propto (-i z)^{-b}$), the
pole becomes a cut at $\omega = \pm (\hat{c}_\infty + i\Gamma) q^{1-
\frac{b}{2}}$. If we include $\beta$ - relaxation like phenomena with
$\omega m''(\omega) \sim \omega^a$, we obtain in the frequency range,
where this fractal behavior holds, 
two strongly damped
propagating modes at $\omega = \pm c_\infty q + i \Gamma q^{1+a}$.
Since the fractal behavior is experimentally observed in depolarized
spectra even below $T_g$, where the $\alpha$
relaxation is far below 
the experimental frequency range 
(see e.g. \cite{surovtsev98}), it could account for the anomalous
strong damping of the  
Brillouin line in the sub $T_g$ regime \cite{surovtsev98}. In this
context it is important to note, that the actual physical reason for
the appearance of a fractal part of the susceptibility spectrum is not
relevant for the described mathematical mechanism of producing strong
damping of phonon modes. The $\beta$ - relaxation phenomenon close to
the critical temperature of mode coupling theory is as good a
candidate as the still not yet understood fractal behaviour below
$T_g$ (\cite{surovtsev98,ruocco99,ruocco00}). 
For $\tau = 10^{5} \gg  (c_{\parallel} q)^{-1} $ the light
scattering probes a solid with a well defined, now renormalized,
phonon mode with a renormalized sound velocity 
which is given by Eq. (\ref{eq:longsol}).
As discussed above, the speed of sound in the glass is always 
bigger than the speed of sound of the liquid.

The next figure \ref{fig:2} shows the spectral function of
$S_{22}^1(q,z)$, which gives the only contribution 
in forward scattering direction and zero contribution in
backscattering geometry for the depolarized
geometry due to the
$\cos^2 (\Theta/2)$ factor (see Eq.(\ref{eq:hv})). 
The $S_{22}^1(q,z)$ component is strongly
influenced by the coupling of transverse currents to the $m=1$
rotational current. 
\begin{equation}
(\hat{Q} {\cal L} \{j^R_0\}_{21}(q)| \frac{-1}{ z- \hat{Q}{\cal L} \hat{Q}}|
\hat{Q} {\cal L} \{j^T_{\pm 1}\}_{00}(q) ) \approx -\frac{q
K_{SR}\tau}{z\tau+i} 
\end{equation}
Deep in the liquid for $\tau = 1 \ll (c_{\parallel} q)^{-1} $ the
purely diffusive transverse
currents cause the Rytov dip at zero frequencies. This is shown in the inset of
Fig. \ref{fig:2} 
When super-cooling the liquid the Rytov dip disappears
and a broad central peak shows up for $\tau = 100 \approx
(c_{\parallel} q)^{-1} $ which develops a shoulder at the frequency of shear
waves. When further super-cooling the liquid for  $\tau = 10^{5} \gg
(c_{\parallel} q)^{-1} $ the propagating transversal phonon modes of
the solid at $ \pm 
c_{\perp} q$ shows up (see Eq. (\ref{eq:cperp})).

A further contribution to the spectrum of depolarized light scattering
according to Eq. (\ref{eq:hv}) is
the $m=2$ component. Due to the  factor $\sin^2
(\Theta/2)$  it is the only contribution in backscattering geometry. 
Since it does not couple to any hydrodynamic
mode, it has the simple form
\begin{equation}
{S''}^2_{22}(\omega) = S^m_{22}(q=0) \frac{\omega^2_R (\frac{K_R
\tau_R}{(\omega 
\tau)^2 +1} + \nu_R)}{(\omega^2 - \omega_R^2 - \frac{\omega^2 \tau K_R
\tau_R}{(\omega \tau)^2 +1})^2 + \omega^2 (\nu_R + \frac{K_R \tau_R}{(\omega
\tau)^2 +1})^2}
\end{equation}    
  The spectral function of $S_{22}^2(q,z)$ 
in  Fig. \ref{fig:3} shows for  
$\tau = 1 \ll (c_{\kappa} q)^{-1} $ no structure for low energies (far
below the rotation spectra). It has no Rytov dip since there is no
matrix element which couples shear waves to the $m=2$ rotational
currents. The absence of a Rytov dip is shown explicitly in the inset
of Fig. \ref{fig:3}. When super-cooling the liquid a broad central peak
shows up for $\tau = 100 \approx (c_{\kappa} q)^{-1} $ which narrows in
the solid for $\tau = 10^{5} \gg (c_{\kappa} q)^{-1} $. In our units the
height of the peak is always $K_R \tau + \nu_R$ and the width is of order
$1/\tau$. 
The absence of transverse modes in backscattering
geometry is clearly seen in experiments (see e.g. \cite{comez99}  or
earlier \cite{li92} where light scattering in
backscattering geometry was applied to the molecular glass former
salol). 
In other scattering geometries the VH - spectrum exhibits ideally only
the transverse 
phonon (for $\omega \tau_S \gg 1$). We want to point out that there
are two qualitatively different mechanism which lead to the observation of
a phonon line in a depolarized light scattering geometry. The first one
is leakage of the longitudinal phonon mode due to an imperfect
polarization filter and should not be present in an ideal depolarized
geometry. The second one at a lower frequency is a direct transverse
phonon which couples to the $S_{22}^1(q,\omega)$ component. 

A further contribution which we have plotted is the $S_{22}^0(q,\omega)$
component. It enters according to Eq. (\ref{eq:vv}) into the intensity
for the polarized light scattering geometry. There is a non zero
matrix element which couples the longitudinal phonon mode to the m=0
rotational mode which we approximate from Eq. (\ref{eq:matrix1}) as
\begin{equation} \label{coupling}
(\{j^T_0\}_{00}(q) {\cal L} \hat{Q} (\hat{Q} (z-{\cal L}) \hat{Q})^{-1}
\hat{Q} {\cal L} \{j^R_{0}\}_{20}(q) ) \approx -\frac{q
K_{lR}\tau}{z\tau+i} 
\end{equation}
Therefore the longitudinal phonon mode shows up in the $m=0$ component
when the liquid is supercooled. This is shown in
Fig. \ref{fig:220}. For $\tau=1$ no coupling to the longitudinal phonon
mode can be observed. This can easily be understood. The coupling term
(\ref{coupling}) reduces to $i K_{lR} \tau q$ for $c_\parallel q \tau \ll 1$
at the position of the phonon mode $\omega =c_\parallel q$.  Since the
coupling term to the phonon mode will appear in second order
perturbation theory in $q$, the phonon mode is multiplied with a factor
$(K_{lR} \tau q)^2$ i.e. the maximum height of the phonon contribution in the
spectrum ${S''}_{22}^0$ will be of order $K^2_{lR} \tau /K_l$, which
is equal 1, in the units we are using in Fig. \ref{fig:220}. 
For $c_{\parallel} q \tau \gg 1$ the coupling term (\ref{coupling}) is
$K_l q/\omega$ i.e. at $\omega=c_\parallel q$,  the coupling constant
is of order 1 and the height of the phonon mode is again of order
$K^2_{lR} \tau /K_l \gg 1$. The phonon mode can be detected as soon as 
$c_{\parallel} q \tau  \sim 1$. In our units this happens  for $\tau \sim
50$.  In Fig \ref{fig:220} a broad shoulder can be seen for $\tau=100$,
which turns into a clearly defined phonon mode ($\tau=10^3,
\tau=10^{5}$). 

The ${S''}_{00}^0(q,\omega)$, ${S''}_{22}^0(q,\omega)$ and
${S''}_{22}^2(q,\omega)$  sum up
according to Eq. (\ref{eq:vv}) to give the total polarized
light-scattering intensity. This is plotted in Fig. \ref{fig:VV} where
we have plotted a fictitious VV-spectrum under the additional
assumption that the squared isotropic part of the polarizability is
ten times as big as the anisotropic one ($a^2 = 10 g^2$). 
Note that the orientational correlator $S_{22}^m(q,\omega)$ and the
translational center of mass component $S_{00}^0(q,\omega)$ are of
completely different origin even though a mixing of the poles occurs
in the supercooled regime. This is best seen in the liquid regime
where the microscopic frequency of the orientation is a rotational
motion whereas the microscopic frequency of the center of mass
component is given by the longitudinal phonon mode. 
It is experimentally impossible to extract
$S_{00}^0(q,\omega)$ out of a measured spectrum except for the case of
vanishing anisotropy.

\subsection{Susceptibility spectra}
 
\begin{figure}
\unitlength1cm
\epsfxsize=10cm
\begin{picture}(9,8)
\put(-0.5,-1){\rotate[r]{\epsffile{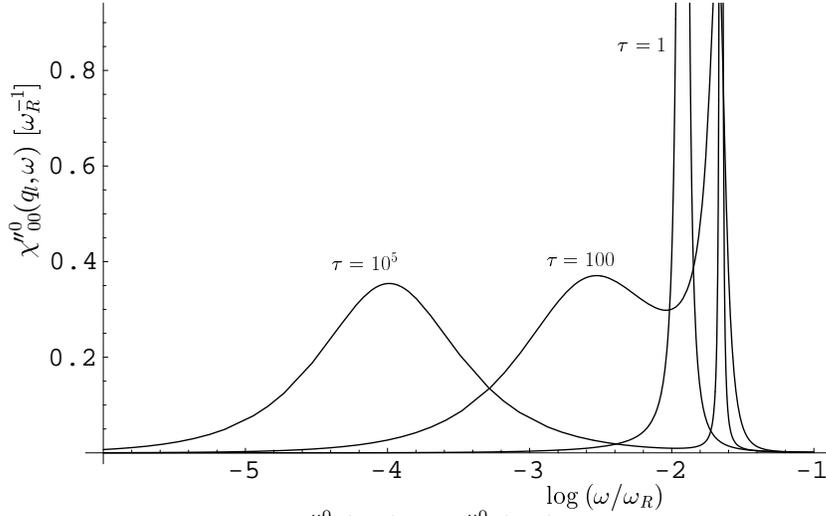}}}
\end{picture}
\caption{The susceptibility  spectrum $ {\chi''}_{00}^0(q,\omega)
= \omega {S''}_{00}^0(q,\omega)$ for the same parameters as in
Fig. \protect\ref{fig:dep} on a logarithmic frequency
scale. 
}
\label{fig:4}
\end{figure}

\begin{figure}
\unitlength1cm
\epsfxsize=10cm
\begin{picture}(9,8)
\put(-0.5,-1){\rotate[r]{\epsffile{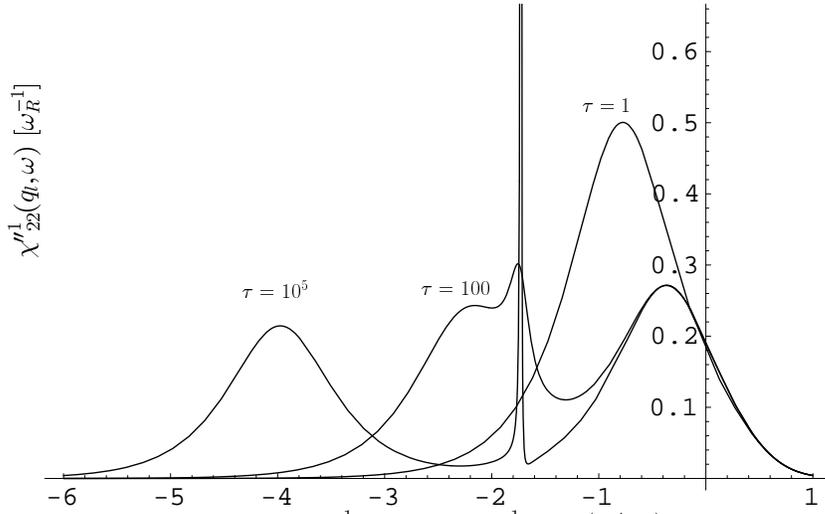}}}
\end{picture}
\caption{The susceptibility spectrum ${\chi''}_{22}^1(q,\omega) =
\omega 
{S''}_{22}^1(q,\omega)$ on a logarithmic frequency
scale. The occurrence of the transverse sound wave can be clearly seen
in the supercooled regime.
}
\label{fig:5}
\end{figure}

\begin{figure}
\unitlength1cm
\epsfxsize=10cm
\begin{picture}(9,8)
\put(-0.5,-1){\rotate[r]{\epsffile{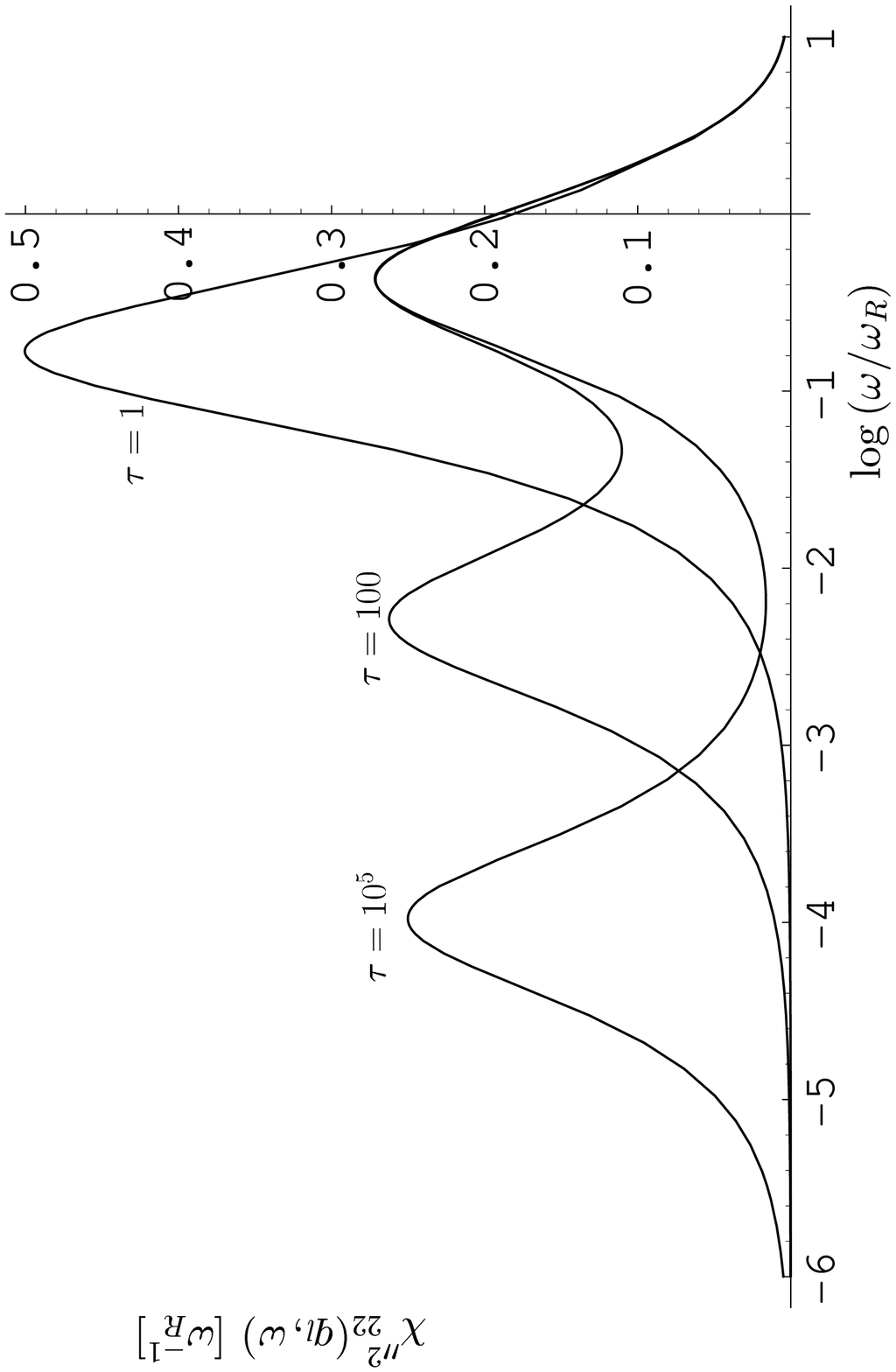}}}
\end{picture}
\caption{The susceptibility spectrum ${\chi''}_{22}(q,2,\omega) =
\omega  {S''}_{22}^2(q,\omega)$ As in Fig. \protect\ref{fig:1} on a
logarithmic frequency scale.
}
\label{fig:6}
\end{figure}

\begin{figure}
\unitlength1cm
\epsfxsize=10cm
\begin{picture}(9,8)
\put(-0.5,-1){\rotate[r]{\epsffile{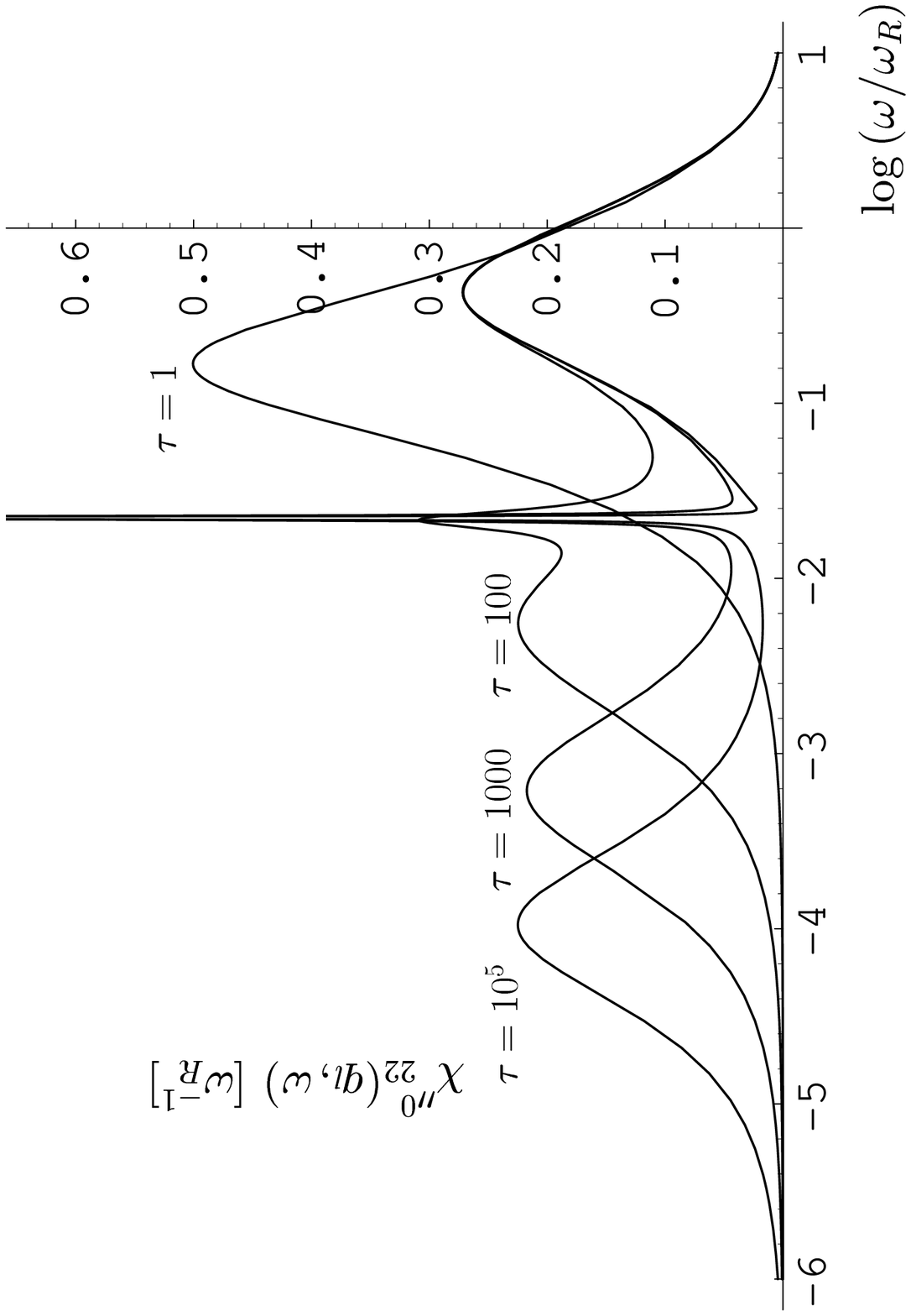}}}
\end{picture}
\caption{The susceptibility spectrum ${\chi''}_{22}^0(q,\omega) =
\omega  {S''}_{22}^0(q,\omega)$ As in  Fig. \protect\ref{fig:220} on a
logarithmic frequency scale. 
}
\label{fig:220log}
\end{figure}

\begin{figure}
\unitlength1cm
\epsfxsize=10cm
\begin{picture}(9,8)
\put(-0.5,-1){\rotate[r]{\epsffile{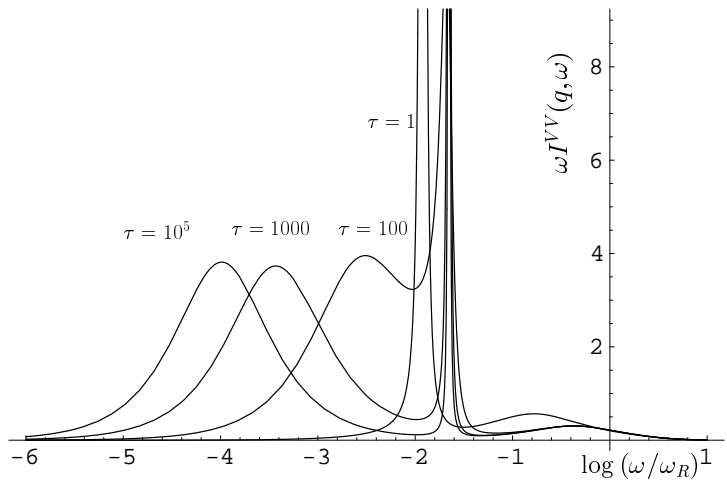}}}
\end{picture}
\caption{The susceptibility spectrum  $\omega I^{VV}(q,\omega)$
on a logarithmic frequency scale. 
}
\label{fig:VVlog}
\end{figure}

In order to make  the influence of structural relaxations 
more transparent, we have plotted on a logarithmic
frequency scale the spectral 
functions of the susceptibilities
\begin{equation}
{\chi''}_{ll'}^m(q,\omega) = \omega {S''}_{ll'}^m(q,\omega)
\end{equation}
which correspond to the quantities plotted in Figs. \ref{fig:1},
\ref{fig:2}, \ref{fig:3}, \ref{fig:220} and \ref{fig:VV}. 
Fig. \ref{fig:4} shows the spectrum  
$ {\chi''}_{00}^0(q,\omega)$. The central peak of Fig. \ref{fig:1} turns into
an $\alpha$--peak which upon super-cooling the glass transition moves
out of the microscopic frequency which is given by the longitudinal
phonon mode. We want to stress again that due to our particular simple
choice of the memory function (Maxwell
theory) the
$\alpha$-peak shown in
Figs.\ref{fig:4},\ref{fig:5},\ref{fig:6},\ref{fig:220log},\ref{fig:VVlog}
does not have the correct stretched form known for glassy
systems and there is no true $\beta$ relaxation regime. 
If the molecular system has no anisotropy (in Eq. (\ref{eq:vv})
$g=0$) this contribution to light scattering is the only direct one.
The area under the $\alpha$-peak $I_{\alpha}$ on a logarithmic
frequency scale, which is also the area under
the Mountain peak \cite{mountain66} on a linear frequency scale 
and the overall intensity $I_{tot}$ are in this case related to the
non-ergodicity 
parameter (Edwards--Anderson parameter)
$f_{00}^0(q,\omega) = I_{\alpha}/ I_{tot}$ for the center of mass component.

The figure (Fig. \ref{fig:5}) shows the spectral function of 
$ \chi_{22}^1(q,z)$. The microscopic is given by the rotational $l=2$
mode roughly at $\omega_R$. When the liquid is supercooled towards the
glass transition the shear wave shows up when the $\alpha$--relaxation
moves over the frequency range range for transverse phonons at $\omega
= c_{\perp} q$. As expected the $m=2$ component $ {\chi''}_{22}^2(q,\omega)$ in
Fig. \ref{fig:6} only shows the broad rotational mode and the
$\alpha$-relaxation since there is no matrix element in the memory
kernel which couples to the $m=2$ component.

Similar to the $m=1$ component the $m=0$ component also shows a
coupling of a hydrodynamic mode. However in this case it is the
longitudinal phonon which couples into the $m=0$ susceptibility. The
spectral function of $ \chi_{22}^0(q,z)$ on a logarithmic frequency
scale is plotted in Fig. \ref{fig:220log}.

The overall situation for a fictitious spectrum $\omega
I^{VV}(q,\omega)$ is shown in Fig. \ref{fig:VVlog} for the same
parameters as in Fig. \ref{fig:VV}. Note that due to
the mixture of $l=l'=0$ and $l=l'=2$ components the Brillouin line of
the longitudinal phonon consists of two components. One is caused by
the direct observation of the center of mass component where the
longitudinal phonon gives a low lying microscopic frequency whereas
the other contribution comes from the phonon mode coupling into the
$l=l'=2$, $m=0$ component. 

\subsection{Light-scattering near an orientational instability} 

Further physics which is contained already in Eq. (\ref{eq:matrix1}) are
some aspects of
light scattering near an orientational instability (e.g. near an
isotropic--nematic transition). From Eq. (\ref{eq:matrix1}) one can see
that the amplitude of the light scattering intensity is for the
depolarized light scattering spectra mainly given by the static
density correlation $S_{22}^m(q)$. Close to a nematic transition, a
weakly first order phase transition, $S_{22}^m(q)$ increases strongly
for small $q$ where $\lim_{q \rightarrow 0}   S_{22}^m(q) ^{-1} =
\kappa_{Kerr}$ is the optical Kerr constant. On the other hand the
rotational frequency $\omega^R$ (in Eq. (\ref{eq:rot})) contains
$S_{22}^m(q)^{1/2}$ in the denominator. In this way our equations
describe the broad central peak together with the strong scattering
intensities at the isotropic--nematic transition.     

\section{Relation to other theories} \label{sec:other}
The main issue of the paper so far, was to demonstrate that the theory 
of light scattering can be brought in a form, which is accessible to
tested theories for the dynamics of supercooled molecular liquids and
to show that the structure of the equations of motion reproduce the
light scattering experiments for molecular liquids. We now will show
how the phenomenological equations underlying the most recent theory
of light scattering for molecular liquids \cite{dreyfus99} can be
rigorously derived within our theory. To keep the derivation as simple 
as possible, we will restrict ourselves to the variables
$\rho_{00}, \rho_{2m}, \{j_\mu^T\}_{00}, \{j_0^R\}_{2m}$ used in 
chapter \ref{sec:glasd}. We also will comment on \cite{wang86} and on
\cite{franosch00}.
In appendix \ref{sec:proj} we demonstrate for the theory of Anderson
and Pecora \cite{anderson71} how light scattering theories for linear
molecules based on
projection operator formalisms can be related to our theory.

Dreyfus et. al \cite{dreyfus99} start by writing the continuity equations for 
the center
of mass density fluctuations and the center of mass momentum density
fluctuations, 
\begin{eqnarray}\label{eq:other1}
\frac{\partial}{\partial t} \rho_{00}(q,t) &=& i
\vec{q}\vec{j}(q,t)\\
\frac{\partial}{\partial t} j_i(q,t) &=& 
i q_j \sigma_{ji}(q,t) \;,
\end{eqnarray}

where $j_i$ are the Cartesian components of the center of mass
current fluctuations and $\sigma_{ij}$ are the Cartesian components of 
the stress-fluctuations. To obtain a closed set of equations it is
necessary to write down constitutive equations for the
stress-fluctuations which relate them to the current and density
fluctuations. Instead of using phenomenological ansatzes, as it was
mostly done in the existing literature, we will write down exact
equations for the stress-tensor fluctuations using generalized
constitutive equations. By using  a formalism introduced  in
\cite{latz89} we can express the stress-fluctuations for vanishing
amplitude of the wave-vector exactly by the
fluctuations of the basic set of variables, which we used in section
\ref{sec:glasd}. (For simplicity it is more convenient to use the
Cartesian components of the center of mass current fluctuations) 
\begin{eqnarray}\label{eq:other2}
\sigma_{ij} &=&   \rho_{00}(\vec{q},t) \frac{1}{S_{00}^0(q=0)}
\; (\rho_{00}|\tau_{ij})\nonumber \\ 
&+& i j_r(\vec{q},t) q_k \otimes \frac{m}{N k_BT}\;
(\sigma_{rk}|R'(t)|\sigma_{ij}) 
\\
&+& i \{j_0^R\}_{2m}(\vec{q},t) \otimes \frac{\Theta}{N k_B T} \;  ({\cal L}
\{j_0^R\}_{2m}|R'(t)|\sigma_{ij}) \nonumber
\end{eqnarray}   
Here, repeated indices are summed over and  we defined $A \otimes B =
\int_0^t dt'\; A(t') B(t-t')$.  $R'(t)$ is again the reduced time
propagator acting in the space perpendicular to the chosen density and
current fluctuations.  Therefore no terms of the form $({\cal L}
\rho_{lm}|R'(t)| \sigma_{ij})$ appear in equation
(\ref{eq:other2}). There are in principal terms proportional to the
fluctuations $\rho_{2m}$. But since in Eq. (\ref{eq:other1}) only the 
combination $q_j \sigma_{ji}$ appears, the proportionality factor
contains 
the term $ q_j (\rho_{2m}|\sigma_{ji}) = (\rho_{2m}|{\cal L} j_i) =
({\cal L} \rho_{2m}|j_i) = \sqrt{6} \; (\{j_0^R\}_{2m}|j_i) =0$
 \cite{schilling97}.   Therefore no fluctuations proportional to
$\rho_{2m}$ do contribute to the generalized hydrodynamic equations.
For $q \to 0$ the equations (\ref{eq:other2}) considerably simplify. 
The only symmetric tensor of rank four $(\sigma_{rk}|R'(t)|\sigma_{ji})$, 
which does not vanish for $q \to 0$ is
$(\sigma_{rk}|R'(t)|\sigma_{ji}) = (p|R'(t)|p) \delta_{kr} \delta_{ij} 
+ (\sigma_{ij}^s|R'(t)|\sigma_{ij}^s) (\delta_{rj} \delta_{ki}
+ \delta_{ri}\delta_{kj} - \frac{2}{3} \delta_{ij} \delta_{rk})$,
where $p = \frac{1}{3} \sum_i \sigma_{ii}$ is the scalar part and
$\sigma_{ij}^s$ the  traceless part of the stress-tensor fluctuations.
The third row of equations (\ref{eq:other2}) can be evaluated by
transforming $\sigma_{ij}$ to spherical components
$\sigma_{lm}$. Then, for $q \to 0$, the identity 
$\{j_0^R\}_{2m}|R'(t)|\sigma_{lm'}) = \delta_{l2} \delta_{mm'} ({\cal L}
\{j_0^R\}_{20}|R'(t)|\sigma_{20})$ and $\dot{\rho_{2m}}(t) = \sqrt{6}
\{j_0^R\}_{2m}(t)$ can be used. Therefore the
constitutive equations (\ref{eq:other2}) reduce to 
\begin{eqnarray} \label{eq:other3}
\sigma_{ij} &=&   \delta P(t) \delta_{ij}+  \frac{\eta_B(t)}{n} 
\otimes \; i \vec{q}\vec{j}(t) \delta_{ij}\\ 
&+& \frac{1}{n} \eta_S(t) \otimes \tau_{ij}(t) 
\\
&-& \mu(t) \otimes 
\dot{Q}_{ij}(t)\nonumber
\end{eqnarray}   
with $\eta_B(t) =
\frac{1}{k_B T \; V} (p|R'(t)|p)$, $\eta_S(t) =
\frac{1}{k_B T \; V} (\sigma_{ij}^s|R'(t)|\sigma_{ij}^s)$ 
being the generalized bulk viscosity and shear viscosity,
respectively \cite{latz89}. The tensor $\tau_{ij} = i(q_i j_j + q_j
j_i - \frac{2}{3} \vec{q}\vec{j} \delta_{ij})$  is the strain
tensor. The fluctuations $\delta P(t)$ of the internal
hydrostatic pressure due to density fluctuations are given by $K
\rho_{00}(q,t)$, where $K = \frac{k_B T}{S_{00}^0(q=0)}$ is the
(static) bulk module, i.e. the inverse of the compressibility
$\kappa$. The tensor $Q$ is given by 

\begin{equation} \label{eq:other4}
\frac{1}{\sqrt{6}} \left( \begin{array}{ccc}
-\frac{\rho_{20}}{\sqrt{6}} + \frac{\rho_{22}}{2} + 
        \frac{\rho_{2\underline{2}}}{2}& 
-i \frac{\rho_{22}}{2} + i \frac{\rho_{2\underline{2}}}{2}&
\frac{-\rho_{21}}{2} +  
        \frac{\rho_{2\underline{1}}}{2}\\
-i \frac{\rho_{22}}{2} + i \frac{\rho_{2\underline{2}}}{2}&
-\frac{\rho_{20}}{\sqrt{6}} - \frac{\rho_{22}}{2} -  
        \frac{\rho_{2\underline{2}}}{2}& i
\frac{\rho_{21}}{2} + i \frac{\rho_{2\underline{1}}}{2}\\ 

-\frac{\rho_{21}}{2} + 
        \frac{\rho_{2\underline{1}}}{2}& i
\frac{\rho_{21}}{2} + i \frac{\rho_{2\underline{1}}}{2}&  
\sqrt{\frac{2}{3}} \rho_{20}
\end{array} \right)
\end{equation}
 and the function  $\mu(t)$ is the the matrix element $\mu(t) = 
\frac{-\theta}{k_B T N} ({\cal 
L} \{j_0^R\}_{20}|R'(t)| \tau_{20})$. Now we only need another
constitutive equation for the tensor $Q_{ij}(t)$. Using the same
strategy as in the derivation of the constitutive equations for the
stress tensor we  easily derive 
\begin{equation} \label{eq:other5}
\frac{d^2}{dt^2} Q_{ij}(t) = - \omega_R^2 Q_{ij}(t) + \mu(t) \otimes
\tau_{ij}(t) - \{M_{00}^{RR}\}_{22}^{00}(t) \otimes \dot{Q}_{ij}(t)
\end{equation}

Here the same function $\mu(t)$ as in Eq. (\ref{eq:other3}) appears
naturally within the formalism, confirming the Onsager principle. The 
memory-function $\{M_{00}^{RR}\}_{22}^{00}(t)$ is the same as used in
section \ref{sec:mmct}. 
Eqs. (\ref{eq:other1}) - (\ref{eq:other5}) are exactly the equations
used in \cite{dreyfus99}. With our formalism, we can identify 
the phenomenologically introduced functions $\mu(t)$ and $Q_{ij}(t)$ 
of \cite{dreyfus99}.  We also want to emphasize, that the convolution
integrals in time i.e. the retardation effects are a necessary
consequence of the slowing down of structural relaxations and its
effect on the frequency dependent viscosities and the rotation - stress
coupling function $\mu(t)$. The ansatz of Quentrec \cite{quentrec76},
where the viscosities and $\mu(t)$ are replaced by only temperature
dependent functions is therefore not acceptable for the description of 
supercooled liquids. 

It is important to note, that the form of the equations
(\ref{eq:other1}) - (\ref{eq:other2}) depends crucially on the chosen
set of variables. If we would {\em not} have chosen the rotational currents
$\{j_0^R\}_{2m}(\vec{q},t)$ explicitly as a member of our basic set of 
equations the last line of Eq. (\ref{eq:other3}) would contain a
coupling $\hat{\mu}(t) \otimes Q_{ij}(t)$ to the tensor $Q_{ij}$
instead of to its time derivative. The function $\hat{\mu}(t)$ can
also be expressed in terms of a (modified) reduced time propagator
$\hat{R}(t)$, $\hat{\mu}(t) \propto 
(\{j_0^R\}_{20}|\hat{R}'(t)|\tau_{20})$. In addition, the equation for
$Q_{ij}(t)$ were of first order in time instead of second order. Wang
uses a mixed representation \cite{wang86}. His constitutive equation
for the stress tensor  coincides with Eq. (\ref{eq:other2}), 
but the equation for $Q_{ij}$ is only of first order. For deriving
exactly such a set of equations, it were necessary to use different
projection operators for deriving the constitutive equation and
the equation for the tensor $Q{ij}$.  
From our point of view there are mainly two reasons why it seems  more 
advantageous to choose one single set of
basis variables including the currents 
$\{j_0^R\}_{2m}(\vec{q},t)$. First,  
since approximation schemes for force - force autocorrelation
functions i.e. the memory-function  $\{M_{\mu\mu'}^{\alpha
\alpha'}\}_{ll'}^{mm'}(t)$  seem to be easier, than for mixed current
- force memory-functions, which would appear, when only the densities
$\rho_{lm}(\vec{q},t)$ and the conserved currents
$\{j_\mu^T\}_{lm}(\vec{q},t)$ were used as variables. Second, by using 
one set of variables the Onsager relations are automatically
fulfilled, since there appears the same function $\mu(t)$ in the
equation for the stress and the tensor $Q_{ij}(t)$. 
In the approach of Wang it is important to choose the
approximations for the different functions $\hat{\mu}(t)$ and
$\mu(t)$ carefully, such that the Onsager principle is guaranteed. (Essentially
the time derivative of $\hat{\mu}(t)$ is related to $\mu(t)$.)  

The approach of  \cite{franosch00} is more general than ours, since
no  assumptions on the scattering mechanisms were used to derive the
general form of light scattering spectra of supercooled liquids.  This 
was achieved by only using the hydrodynamic variables center of mass
density and center of mass currents as the basic set of variables for
applying constitutive equations to the dielectric tensor
fluctuations. In this way the coupling mechanisms between rotation and
translations in molecular liquids are not explicitly treated but lead 
implicitly to frequency dependent Pockels constants, relating the
hydrodynamic modes 
to the dielectric tensor fluctuations, and unknown background
spectra. Due to the general nature of the approach in principle also
other mechanisms like 
DID are contained on a formal level in the description, although
they cannot be explicitly treated without using a specific theory.  

In our approach we concentrated on a specific mechanism for the
coupling of light to the motion of the linear molecules, by assuming
that the  principal axis of the polarizability tensor agree with the
principal axis of the inertia tensor of the molecule. Under this
assumption Eq. (\ref{eq:vv}) and (\ref{eq:hv}) are completely general.
The importance of 
the index of helicity $m$ seems to be at variance with the approach of 
\cite{franosch00}, since there only $m=0$ components appear. But if 
we express $S_{22}^m(q,t)$ by correlation functions of the hydrodynamic
variables plus a background spectra with the help of the exact generalized
constitutive relations as used above  we obtain in the $q$ -
frame for $q\to 0$

\begin{eqnarray}\label{hydroform}
S_{22}^m(z) &=& (\rho_{2m}|{R'}^{NH}(z)|\rho_{2m}) \nonumber \\
&&+ q^2 
((\rho_{2m}|{R'}^{NH}(z)|\sigma_{zz})\frac{m}{k_B T})^2
(j_\parallel|R(z)|j_\parallel)\\
&&  +  2 q^2 
((\rho_{2m}|{R'}^{NH}(z)|\sigma_{xz})\frac{m}{k_B T})^2
(j_\perp|R(z)|j_\perp) \nonumber
\end{eqnarray}  

Here the reduced resolvent ${R'}^{NH}(z) = \frac{-Q}{z - Q {\cal L}Q
  }$ describes  
the dynamic perpendicular to the hydrodynamic fluctuations only
i.e. the subscript $NH$ indicates, that the projection operator $Q$,
used in Eq. (\ref{hydroform})  projects on the space of
non hydrodynamic variables.  
Transforming $\sigma_{ij}$ to irreducible  spherical components, the
$I^{VH}$  spectrum (\ref{eq:hv}) can be written
\begin{equation} \label{frano}
I^{VH}(z) = {\cal T}(z) - q^2 \cos^2(\Theta/2) a_{VH}^2(z)
(j_\perp|R(z)|j_\perp) 
\end{equation}
where  the background spectrum ${\cal T}(z)$ and the generalized
Pockels constant $a_{VH}(z)$ are given by 
\begin{eqnarray} \label{pockels}
{\cal T}(z) &=& g^2 \frac{4 \pi}{15} (\rho_{20}|{R'}^{NH}(z)|\rho_{20})
\nonumber\\ 
 a_{VH}(z)&=& g \sqrt{\frac{4 \pi}{15}}
(\rho_{20}|{R'}^{NH}(z)|\sigma_{20})\frac{m}{k_B T} 
\end{eqnarray}

In leading order in q (i.e. $q \equiv 0$) we are now allowed to
replace the reduced resolvent in
Eq. (\ref{pockels}) with the full resolvent \cite{franosch00}. 
Similar equations can be derived for the $I^{VV}$ spectrum. 
Eq. (\ref{frano}) is exactly the form found in \cite{franosch00}. The
explicit $m$ dependence is not present anymore in (\ref{frano}), since 
we used that at $q=0$ correlation functions between fluctuations with
the same $l$ and $m$, which do not contain
hydrodynamic poles, do not depend on $m$ anymore. But it is important
to note, that the dynamic coupling to the transversal current
fluctuations is not vanishing, only due to the existence of an
irreducible $m=1$
component of the stress tensor. Also in \cite{franosch00} this
symmetry was implicitly used. But after using it, 
the reduced matrix element $(\rho_{21}|{R'}^{NH}(z)|\sigma_{21})$ 
can be replaced 
by the one for $m=0$ at $q=0$ in Eq. (\ref{hydroform}).  
The $m$ dependence is replaced by an explicit dependence on the 
transversal current
spectrum, which is only present due to the coupling of 
$S_{22}^1(q,z)$ to the transversal current fluctuations.  We could now
further proceed and express the background spectrum and the Pockels
constant with the  method, explained in the appendices \ref{sec:proj} -
\ref{appendixa}, 
 by the correlation functions and memory functions,
which we used in sections \ref{sec:mmct} and \ref{sec:glasd}. But
since we have to evaluate the center of mass current correlation
functions and the center of mass density correlation function at small
but finite values of q (for example to be able to understand the
renormalization of the transversal sound waves by rotation translation 
coupling), we would arrive
at exactly the same theory, which we already derived in
the mentioned sections. The merit of the approach of \cite{franosch00}
is to work out the most general form of light scattering spectra using
only generalized hydrodynamic and generalized constitutive equations
for dielectric fluctuations. Thus rigorous constraints concerning the
appearance of hydrodynamic excitations in  different scattering 
geometries  are formulated. 
For explicit considerations of specific scattering mechanisms, as done
in this paper,  one has in general to include also non hydrodynamic
variables.

\section{Conclusion}

In this work we developed  a microscopic theory of light scattering for linear
molecules,  concentrating on the direct contribution
to the spectra. The starting point of our theory  is an
exact expression for the spectra in terms of correlation functions
(Eqs. (\ref{eq:vv})  and
(\ref{eq:hv})). It turns out to be important to accurately
take the tensorial character of the orientational correlation function
into account.  
This is due to the fact that the orientational components
of different helicity index $m$ transform in general
differently under rotations.  It is in this context crucial, that the
dynamic correlation functions $S_{22}^m(q,\omega)$ (contrary to the
memory functions) have to be
evaluated at small but {\em
finite} wave vectors due to the following reason. The 
rotational symmetry allows for the dynamic coupling  of $m=0$ components of
tensorial densities $\rho_{lm}$ and 
rotational currents to the
longitudinal and the $m=1$ components of
tensorial densities $\rho_{lm}$ and rotational currents 
to transversal current
fluctuations, respectively. Microscopically the coupling is
non vanishing due
to the induction of local stress by the rotation of the
molecules.  Therefore the hydrodynamic poles show up in the
corresponding dynamic correlation function $S_{22}^m(q,\omega)$. 
Thus we are not allowed to replace them by there value at $q=0$, where
indeed correlation functions for different values of $m$, but the same
$l$, are equal. Only the $m=2$ component of $S_{22}^m(q,\omega)$ which
does not couple to any hydrodynamic mode and
all memory functions, which -- due to our choice of dynamic variables --
do not contain by construction any
hydrodynamic pole, 
can be replaced by its value at $q=0$. It is
the violation of rotational symmetry on the spatial scale of the light
scattering experiments due to the existence of hydrodynamic modes,
which causes the importance of  the helicity index $m$. 

Based on a projection operator formalism, we formulate a microscopic
theory for the correlation functions $S_{ll'}^m(q,\omega)$ of
supercooled molecular liquids, which include all possible
couplings to hydrodynamic modes. Simplifying the equations of the
molecular mode coupling theory extended by transverse currents, we
demonstrate explicitly, that a
qualitative description for light scattering spectra near the
glass transition can be achieved, which treats correctly the interplay
of hydrodynamic modes and structural relaxations. 
We further derive microscopic expressions which give the influence of
the rotation--translation coupling onto the hydrodynamic poles. It is
also shown how other theories of light scattering can be expressed by 
the quantities, which appear in our theory.

The equations (\ref{eq:vv}) and (\ref{eq:hv}) are in principle not
restricted to small q-values. They are therefore also valid for the
interpretation of X-ray spectra, if we give up the restriction to
small wave vectors. Since the MMCT, formulated in our
paper is a theory for all wave vectors, there will be  no principal problems 
to do this. 
A possible application  of the restricted
theory in \ref{sec:glasd}, would be   
to compare
spectra of different scattering angles and of different scattering
geometries in order to obtain microscopically relevant quantities like
e.g. the rotation--translation coupling. 
But to obtain reliable results, it is of course necessary to give up
the Maxwell ansatz and to include $\beta$ relaxation phenomena in the
memory functions. 
Further we like to encourage
the evaluation of the orientational components for different $m$
values from computer simulations.

\acknowledgements
We thank R. Schilling, M. Fuchs, H.~Z.~Cummins and R.~Pick for a
critical reading of the manuscript and for helpful comments.  Our work
was financially supported from the SFB 262. 

\begin{appendix}
\section{Coupling between polarizability and shear stress}
\label{sec:proj}

To demonstrate how theories of
light-scattering, based on various projection operator formalism can in
principal  be expressed by the correlation functions appearing in
molecular MCT, we take as an example the set of variables from
the book
of Berne and Pecora \cite{berne76} and show how their memory functions 
can be evaluated using the basis set of molecular MCT. 

We use the projection operator formalism in Laplace transformed space
(see e.g. \cite{forster83}).
To explain the appearance of a Rytov Dip,  in \cite{berne76} 
the polarizability and one component of the
transverse current $j_x(q,z)$ with $j_x \sim 1/\sqrt{2} (
\{j^T_1\}_{00} + i \{j^T_{-1}\}_{00})$ are chosen as a minimal basis set. 
$z = \omega + i \epsilon$ is the complex frequency.
We have shown in section \ref{sec:sec3} that the polarizability for the
depolarized light scattering is in the subspace of the l=2
density. Therefore we define $\alpha_{VH}(q,z) =: \rho_2 = \sum_m B_m
\rho_{2m}$. 
Applying  Mori--Zwanzig projection technique with a projection
operator $P_{BP} = |\rho_2)\frac{1}{(\rho_2|\rho_2)}(\rho_2| +
|j_x)\frac{1}{(j_x|j_x)}(j_x|$ and 
$Q_{BP}= \underline{\underline{1}} - P_{BP}$, 
the dynamics of the polarizability correlation function for the
depolarized light scattering results from a solution of the following
2x2 matrix equation:
\begin{eqnarray}
\lefteqn{
\left (
\begin{array}{cc}
z-(\rho_2 { \cal L } R_1' { \cal L } \rho_2 ) & -(\rho_2 { \cal L } R_1'
{ \cal L } j_x) \\
-(j_x  {\cal L} R_1' { \cal L } \rho_2 ) &
z - (j_x {\cal L} R_1' {\cal L} j_x )
\end{array}
\right ) \mbox{\hspace*{3cm}}} \nonumber \\ && \mbox{\hspace*{3cm}}
\left ( (\rho_2 \rho_2 )(q,z) \atop 0 \right ) = 
\left ( (\rho_2 \rho_2 )^0 \atop 0 \right )
\label{eq:bp}
\end{eqnarray}
which is still exact. ${\cal L}$ is the Liouvillian and
$R_1' = Q_{BP} ( z - Q_{BP} {\cal L }
Q_{BP})^{-1} Q_{BP} $ is the reduced dynamics and $ (\rho_2 \rho_2
)^0 $ is the static correlation function. The occurrence of a dip in
the spectrum relies on the fact that the off 
diagonal element $( \rho_2 { \cal L } R_1'
{ \cal L } j_x) $   does not vanish for small but finite wave-vector
$q$. It follows from momentum conservation, that it is of order $q$. It
is therefore possible to define an  
effective coupling constant $R$ between transversal current
fluctuations and polarizability fluctuations due to the rotation of
the molecule (see ref. \cite{berne76} p. 317 to have an explicit
connection between the phenomenological theory in \cite{berne76} based
on an incomplete basis set for the projection technique and our
microscopic theory).
\begin{equation} \label{coupling2}
R \sim \lim_{z \rightarrow 0+i\epsilon} \lim_{q \rightarrow 0} \frac{1}{q^2} 
\left | ( \rho_2 { \cal L } R_1'
{ \cal L } j_x) \right |^2 \ne 0
\label{eq:rbp}
\end{equation}

In the following we show, how this matrix element can be expressed by
the memory-functions of MMCT. In this theory not only the density
$\rho_{2m}$, but also the corresponding currents $\{j_0^\alpha\}_{2m}$
are used as additional variables. A coupling of the form
Eq. (\ref{coupling2}) can therefore not appear since the reduced
resolvent $R_2'$ in the new set of variables is projecting perpendicular to
the currents i.e also perpendicular to ${\cal L} \rho_2$ as defined
above.  We have shown in detail in sec. \ref{sec:glasd}:
\begin{eqnarray}
{\cal L} \rho_2 &=& \sum_m B_m {\cal L} \rho_{2m}(q,z) =   \sum_m B_m
q \{j^T_0\}_{2m}(q,z) + \sqrt{2(2+1)} \{j^R_0\}_{2m}(q,z) 
\nonumber \\ 
&=&  
\sum_m B_m \sqrt{2(2+1)} \{j^R_0\}_{2m}(q,z) + O(q) 
\nonumber \\
 &=:&
\label{eq:lr}
j_2
\end{eqnarray}
where $T$ refers to translational currents which occur when applying
the Liouvillian on the time dependent positions and $R$ refers to
rotational currents which appear when applying the Liouvillian to the
time dependent orientations. For small wave-vectors, we can neglect the
contribution of translational currents in the following analysis. 
We also showed in \ref{sec:glasd}, that only the $m=1$ component of the
rotational current ${\cal L} j_2$ has non-vanishing matrix elements
with ${\cal L} j_x$. 
Applying again Mori-Zwanzig projection
technique with the enlarged Hilbert space with
\begin{equation}
P_L = | \rho_2 )(\rho_2 | + | j_x )( j_x | + | j_2 )( j_2 | 
\end{equation}
and $Q_L = \underline{\underline{1}} - P_L$ leaves a 3x3 Matrix
equation to be solved. Note that due to Eq. (\ref{eq:lr})
contributions that contain $Q_L {\cal L} \rho_2$ vanish.
\begin{eqnarray}
\lefteqn{
\left (
\begin{array}{ccc}
z & & -(\rho_2 { \cal L }  j_2) \\
& z -(j_x  {\cal L} R_2' { \cal L } j_x ) & -(j_x  {\cal L} R_2' {
\cal L } j_2 ) \\
-(j_2 { \cal L }  \rho_2) & -(j_2  {\cal L} R_2' {
\cal L } j_x ) &
z - (j_2 {\cal L} R_2' {\cal L} j_2 )
\end{array}
\right ) \mbox{\hspace*{3cm}}} \nonumber \\ && \mbox{\hspace*{3cm}}
\left ( \begin{array}{c}(\rho_2 \rho_2 )(q,z) \\ 0 \\ 0 
\end{array} \right ) = 
\left ( \begin{array}{c} (\rho_2 \rho_2 )^0 \\ 0 \\
 0 \end{array} \right )
\label{eq:lsmct}
\end{eqnarray}
here $R_2'$ is the reduced dynamics due to the new variable set:
\begin{equation}
R_2' = Q_{L} ( z - Q_{L} {\cal L }
Q_{L})^{-1} Q_{L} 
\end{equation}
Making an additional projection step to obtain an
effective $2$x$2$--matrix with $|\rho _2)$ and $|j_x )$ as variables
(see appendix \ref{appendixa}) gives a theory for the matrix elements
of Eq. (\ref{eq:bp}):
\begin{eqnarray}
\left ( \rho_2 {\cal L} R_1' {\cal L} \rho_2 \right ) &=& 
\frac{ \left | \left ( \rho_2 {\cal L} j_2
\right ) \right | ^2}
{z- \left ( j_2 {\cal L} R_2' {\cal L} j_2 \right )}
\nonumber \\
\left ( j_x {\cal L} R_1' {\cal L} j_x \right ) &=&
\left ( j_x {\cal L} R_2' {\cal L} j_x \right )  +
\frac{ \left | \left ( j_x {\cal L} R_2' {\cal L} j_2 \right )
\right | ^2}{z- \left ( j_2 {\cal L} R_2' {\cal L} j_2 \right )}
\nonumber \\
\left ( j_x {\cal L} R_1' {\cal L} \rho_2 \right ) &=&
\frac{ \left ( j_x {\cal L} R_2' {\cal L} j_2 \right )
\left ( j_2 {\cal L} \rho_2 \right ) }
{z- \left ( j_2 {\cal L} R_2' {\cal L} j_2 \right )}
\label{eq:r1eff}
\end{eqnarray} 

The transformation to the complete set of variables used in MMCT involves
inversions of much larger matrices, but the strategy will be the same.
For $q \to 0$ only the matrix elements in Eq. (\ref{eq:r1eff}) should be
 relevant.  
Eq. (\ref{eq:r1eff}) together with the mode coupling expressions in
sec. \ref{sec:mmct},  therefore constitute  a microscopic theory for 
the effective coupling coefficient Eq. (\ref{coupling2}) of the theory
described in \cite{berne76}.  

\section{Connection between different projection schemes}
\label{appendixa}

Lets assume we have a particular basis  system ${A_i,B_k}$ with
$(B_i|A_k) =0$ for all $i,k$,  
which spans the subspace
${\cal H}^L$.  An example for that would be the basis set of MMCT
described in sec. \ref{sec:mmct}, with $A_i$ being the densities
$\rho_{lm}$ and $B_k$ the currents $\{j_\mu^\alpha\}_{lm}$. With such
a basis set  of the Hilbert space a projection operator
$\hat{P}^L$ can be defined, which projects into the subspace ${\cal H}^L$. 
Within the subspace ${\cal H}^L$ we have a theory to calculate the
matrix elements of the memory function. In order to be able to compare
different projection schemes  using a  reduced set of variables
$A_i$,  which are  
elements of a subspace ${\cal H}^{A} \subset {\cal H}^{L}$ we need a formalism
which expresses all matrix elements in the subspace ${\cal H}^{A}$ as
functions of the matrix elements in the subspace ${\cal H}^L$. 
This can be achieved by applying a formalism described in
\cite{latz89}. 

In ${\cal H}^L$ the operator $(z-{\cal L})^{-1}$ is given by:
\begin{eqnarray}
\hat{P}^L (z-{\cal L})^{-1} \hat{P}^L  &=& \left [
\hat{P}^L (z-{\cal L}) \hat{P}^L - 
\hat{P}^L {\cal L} \hat{Q}^L ( \hat{Q}^L (z-{\cal L}) \hat{Q}^L )^{-1}
\hat{Q}^L {\cal L} \hat{P}^L \right ]^{-1} \nonumber \\
&=& \left [ \hat{P}^L (z-{\cal L}) \hat{P}^L - \hat{P}^L {\cal L}
\hat{R}^{L'} {\cal L} \hat{P}^L  \right ]^{-1}
\end{eqnarray}
where $ \hat{Q}^L = \hat{1} - \hat{P}^L $ is the usual  projector
perpendicular to $\hat{P}^L $ and 
$\hat{R}^{L'}$ is the reduced dynamics.  

Due to the orthogonality of the variables $A_i, B_k$ we can decompose
$\hat{P}^L$ in $\hat{P}^L = \hat{P}^A + \hat{P}^B$  such that $\hat{Q}^L +
\hat{P}^B = \hat{1} - \hat{P}^A$.
In the subspace ${\cal H}^A$ we can write down a similar equation as above:
\begin{equation}
\hat{P}^A (z-{\cal L})^{-1} \hat{P}^A = \left [
\hat{P}^A (z-{\cal L}) \hat{P}^A - \hat{P}^A {\cal L}
\hat{R}^{A'} {\cal L} \hat{P}^A \right ]^{-1}
\end{equation} 
where the reduced dynamics is in the subspace ${\cal H}^A$ is given by
\begin{equation}
\hat{R}^{A'} = \hat{Q}^A \left [
\hat{Q}^A (z-{\cal L}) \hat{Q}^A
\right ] ^{-1} 
\hat{Q}^A 
\end{equation}
Let us now express $\hat{R}^{A'}$
depending on the reduced dynamics $\hat{R}^{L'}$. We will therefore
have to do an inversion of the operator $\hat{M}:=(\hat{Q}^A ({\cal
L}-z) \hat{Q}^A)$ 
\begin{equation}
\hat{M} = \left (
\begin{array}{cc}
\hat{P}^B  ({\cal L}-z) \hat{P}^B & 
\hat{P}^B  ({\cal L}-z) \hat{Q}^L  
\\
\hat{Q}^L ({\cal L}-z) \hat{P}^B  & 
\hat{Q}^L ({\cal L}-z) \hat{Q}^L
\end{array} \right )
\end{equation}
The inversion of $\hat{M}$  gives \cite{latz89}:
\begin{equation}
\hat{M}^{-1} = \hat{R}^{A'} = \hat{R}^{L'} - ( \hat{P}^B -
\hat{R}^{L'} \hat{M} ) \hat{K} ( \hat{M} \hat{R}^{L'} - \hat{P}^B)
\end{equation}
with
\begin{equation}
\hat{K} = \left (\hat{P}^B (-z+{\cal L}) \hat{P}^B  -
\hat{P}^B {\cal L} R^{L'} {\cal L} \hat{P}^B
\right ) ^{-1}
\end{equation}
Therefore the connection between the two reduced dynamics $R^{A'}$ and
$R^{L'}$ are given by:
\begin{equation}
\hat{R}^{A'} = \hat{R}^{L'} - ( \hat{P}^B -
\hat{R}^{L'} {\cal L} \hat{P}^B ) \hat{K}  
(\hat{P}^B {\cal L} \hat{R}^{L'} - \hat{P}^B) 
\end{equation}
This connection was used in Eq. (\ref{eq:r1eff}) to derive the
connection between the dynamics in the two different basis sets.

\section{Matrix elements of the polarizability tensor}
\label{app:a}

The polarizability of every molecule Eq. (\ref{alpha}) is a 
tensor of rank 2. It can be written as a scalar plus an irreducible
tensor of rank 2.
In a body fixed coordinate system, with $\hat{n}^i$ chosen along the
principal axis, it has the simple form:
\begin{equation} \label{albody}
\mbox{\boldmath $\alpha$}^B =a \left( \begin{array}{ccc}
 1&&\\&1&\\&&1\\ \end{array} \right)  
+ \frac{2 g}{3} 
\left ( \begin{array}{ccc} -\frac{1}{2}&&\\&-\frac{1}{2}&\\&&1
\end{array} \right ) 
\end{equation}

The irreducible spherical components are calculated by:
\begin{equation} \label{sphericaltrafo}
\alpha_{lm} = \sum_{i,j} \sum_{m_1,m_2} C(11l;m_1m_2m) U_{m_1 i} U_{m_2 j}
\alpha_{ij}^B
\end{equation}

where the $i,j \in \{x,y,z\}$ are Cartesian indices and $m_1,m_2
\in \{-1,0,1\}$  are spherical ``helicity''  indices. $C(l_1l_2l;m_1m_2m)$ 
are the Clebsch Gordan coefficients. The matrix $\bf U$
is given by 
\begin{equation}
U_{m i} =
\left ( \begin{array}{ccc} 
\frac{1}{\sqrt{2}}& -\frac{i}{\sqrt{2}}&0 \\0&0&1
\\-\frac{1}{\sqrt{2}} & -\frac{i}{\sqrt{2}}&0 
\end{array} \right ) 
\end{equation}

In the body fixed frame $\mbox{\boldmath $\alpha$}^B$ only the 
spherical components $\alpha^B_{lm}$ with $m=0$ do not vanish:
$\alpha^B_{00} = - \sqrt{3} \; a$ and $\alpha^B_{20} = \sqrt{\frac{2}{3}}\,
  g$.  The spherical components in the $q$ - frame are easily obtained
by rotation.   
\begin{equation}
\alpha^S_{lm}(\hat{\Omega}_i(t)) = \sum_n D^l_{nm}(\Omega^{-1}_i(t))
\alpha^B_{ln} = D^l_{m0}(\Omega_i(t))^* \equiv \sqrt{\frac{4 \pi}{2l+1}}
Y_{lm}(\Omega_i(t))   
\end{equation}
where $D^l_{nm}(\Omega_i(t))$, $Y_{lm}(\Omega_i(t))$  are
the Wigner
matrices and spherical harmonics, respectively. We used, that the  
angle 
$\hat{\Omega}_i(t)$ denotes the rotation carrying the body fixed frame
into coincidence with the space fixed $q$ - frame. This is the inverse
$\hat{\Omega}_i(t) = \Omega^{-1}(t)$ to
the angle describing the orientation of the molecules with respect to 
the $q$ -  frame.   The Cartesian components
in the $q$ - frame are obtained by applying the inverse transformation 
to (\ref{sphericaltrafo})
\begin{equation} \label{trafo}
\alpha^S_{ij} = \sum_{lm} \sum_{m_1,m_2} U^{-1}_{i m_1}
U^{-1}_{j m_2} C(11l;m_1m_2m) 
\alpha_{lm}^S
\end{equation}

The final result for $\mbox{\boldmath $\alpha$}^S$ is:
\begin{eqnarray} \label{eq:polmat}
&&\mbox{\boldmath $\alpha$}^S(t) = a \; \; \left(\begin{array}{ccc}
1&&\\&1&\\&&1 
\end{array} \right) + \\
&&2 g \sqrt{\frac{2 \pi}{15}} \left( \begin{array}[c]{ccc}
        - \frac{1}{\sqrt{6}} Y_{20}(\Omega(t)) + 
Re(Y_{22}(\Omega(t)))  &
 Im( Y_{22}(\Omega(t)))  &
- Re(Y_{21}(\Omega(t)))  \\ 
 Im (Y_{22}(\Omega(t))) &
 -\frac{1 }{\sqrt{6}} Y_{20}(\Omega(t))  -
 Re(Y_{22}(\Omega(t))) & - 
 Im(Y_{21}(\Omega(t))) \\
-Re(Y_{21}(\Omega(t)))  & -Im(
  Y_{21}(\Omega(t))) & {\sqrt{\frac{2}{3}}}
{Y_{20}(\Omega(t))}
\end{array} \right) \nonumber
\end{eqnarray}
where $Im$, $Re$   denote imaginary and
real part, respectively, 

From the matrix element of Eq. (\ref{eq:polmat}) and
Eqs. (\ref{spectravv}) - (\ref{spectravh}) we can
calculate the contributions which are observable in different
scattering geometries. 
Due to the fact that we use the $q$ - frame as the external coordinate
system the correlation function $S_{ll'}^{mm'}$ are diagonal with
respect to  $m,m'$ and $S_{ll'}^m(q,t) = S_{ll'}^{-m}(q,t)$. With 
\begin{eqnarray*}
\frac{N}{2}
S_{22}^m(q,t) &=& \sum_{i,j}\langle Im
Y_{2m}(\Omega_i(t)) e^{-i \vec{q} \vec{r_i}(t)} Im
Y_{2m}(\Omega_j(0)) e^{i \vec{q} \vec{r_j}(0)} \rangle\\
&=& \sum_{i,j}\langle Re
Y_{2m}(\Omega_i(t)) e^{-i \vec{q} \vec{r_i}(t)} Re
Y_{2m}(\Omega_j(0)) e^{i \vec{q} \vec{r_j}(0)} \rangle   
\end{eqnarray*}
and 
\begin{eqnarray*}
\delta_{m,0}
\frac{N}{2} S_{2l}^0(q,t) &=& \sum_{i,j}\langle Re
Y_{2m}(\Omega_i(t)) e^{-i \vec{q} \vec{r_i}(t)} 
Y_{l0}(\Omega_j(0)) e^{i \vec{q} \vec{r_j}(0)} \rangle 
\end{eqnarray*}
the results (\ref{eq:vv}), (\ref{eq:hv}) are
obtained. 

\end{appendix}


%



\end{document}